\documentclass[english,prb, amsmath, twocolumn]{revtex4-1}
\usepackage{ae,aecompl}
\usepackage[T1]{fontenc}
\usepackage[latin9]{inputenc}
\usepackage{textcomp}
\usepackage{amsmath}
\usepackage{amssymb}
\usepackage{graphicx}
\usepackage{esint}

\makeatletter

\newcommand{\lyxmathsym}[1]{\ifmmode\begingroup\def\b@ld{bold}
  \text{\ifx\math@version\b@ld\bfseries\fi#1}\endgroup\else#1\fi}

\providecommand{\tabularnewline}{\\}

 \@ifundefined{textcolor}{}
 {%
   \definecolor{BLACK}{gray}{0}
   \definecolor{WHITE}{gray}{1}
   \definecolor{RED}{rgb}{1,0,0}
   \definecolor{GREEN}{rgb}{0,1,0}
   \definecolor{BLUE}{rgb}{0,0,1}
   \definecolor{CYAN}{cmyk}{1,0,0,0}
   \definecolor{MAGENTA}{cmyk}{0,1,0,0}
   \definecolor{YELLOW}{cmyk}{0,0,1,0}
 }

\@ifundefined{showcaptionsetup}{}{%
 \PassOptionsToPackage{caption=false}{subfig}}
\usepackage{subfig}
\makeatother

\usepackage{babel}
\begin{document}

\title{Continuum Model of the Twisted Bilayer}

\author{J. M. B. Lopes dos Santos}

\email{corresponding author: jlsantos@fc.up.pt}

\selectlanguage{english}%

\affiliation{CFP and Departamento de F\'{\i}sica e Astronomia, Faculdade de Ci\^{e}ncias,
Universidade do Porto, 4169-007 Porto, Portugal}

\author{N. M. R. Peres}

\affiliation{Graphene Research Center and Department of Physics, National University
of Singapore, 2 Science Dr. 3, Singapore 117542, and Centro de F\'{\i}sica
and Departamento de F\'{\i}sica, Universidade do Minho, P-4710-057,
Braga, Portugal }

\author{A. H. Castro Neto}

\altaffiliation{on leave from Department of Physics, Boston University, 590 Commonwealth Avenue, Boston, MA 02215,USA.}

\selectlanguage{english}%

\affiliation{Graphene Research Center and Department of Physics, National University
of Singapore, 2 Science Dr. 3, Singapore 117542 }
\begin{abstract}
The continuum model of the twisted graphene bilayer\cite{santos-2007}
is extended to include all types of commensurate structures. The essential
ingredient of the model, the Fourier components of the spatially modulated
hopping amplitudes, can be calculated analytically, for any type of
commensurate structures in the low twist angle limit. We show that
the Fourier components that could give rise to a gap in the SE-even
structures discussed by Mele\cite{PhysRevB.81.161405} vanish linearly
with angle, whereas the amplitudes that saturate to finite values,
as $\theta\to0$, ensure that all low angle structures share essentially
the same physics. We extend our previous calculations beyond the validity
of perturbation theory, to discuss the disappearance of Dirac cone
structure at angles below $\theta\lesssim1\text{º. }$ 
\end{abstract}
\maketitle

\section{Introduction}

Barely a year after the discovery of a new form of quantization of
the Hall effect in graphene mono-layers layers\cite{GS05,NMM+06b_short,PGN06b,ZTS+05_short},
the bilayer attracted considerable attention by displaying yet another
type of Quantum Hall effect\cite{NMM+06b_short}. Experimental and
theoretical studies quickly followed, on the electronic structure\cite{OBS+06},
Landau level spectrum\cite{MF06}, transport\cite{Kat06b,KNG06b,KA06,LK06,NNG+06},
disorder and interactions\cite{HLH+06,NNP+06}. 

These early studies focused on the $AB$ stacked bilayer\cite{RevModPhys.81.109}.
Unlike the mono-layer, in which carriers near the Fermi level behave
like massless fermions, the $AB$ stacked bilayer has quadratic dispersion
near the Fermi level (for undoped samples). It is gapless, as the
mono-layer, but only in the absence of a perpendicular electric field.
An important feature of this system is the existence of a variable
energy gap induced by an external electric field perpendicular to
the layers\cite{ISI:000251107500045_short,McC06}. 

The first experimental indications of the existence of rotational
disorder in ultra-thin graphite films came from films grown on the
$\mathsf{4H-SiC(000\bar{1})}$ (Carbon side) of $\mathrm{SiC}$ crystals\cite{hass-2007};
however, it had been known for years that in graphite crystals the
top layer is often found rotated with respect to the underlying ones,
giving rise long wavelength modulations of the STM signals, displaying
as Moiré patterns\cite{RK93,RON94,PD05,PD05b}. Few-layer graphene
films grown by chemical vapor deposition methods\cite{Chen2010,Reina2009,Li2010_short}
often show rotations of successive graphene layers. It has also been
possible to produce twisted bilayers using mechanically exfoliated
samples\cite{ni:235403}.

The electronic structure of the twisted bilayer was first considered
by the authors\cite{santos-2007} in the context of a continuum, Dirac-Weyl
equation, description of the two layers, coupled by a spatially modulated
hopping. The model predicted the persistence of linear dispersion,
with well defined Dirac cones, like in the mono-layer, but with an
angle dependent suppression of the Fermi velocity; it was also predicted
that there would be no gap in the presence of a perpendicular electric
field. These results were subsequently confirmed experimentally by
Raman\cite{ni:235403} and Landau level spectroscopy\cite{PhysRevLett.106.126802},
and by band structure calculations\cite{PhysRevB.81.165105,PhysRevB.82.121407},
although the earliest calculations appeared to question the suppression
of the Fermi velocity\cite{Shallcross2008,Latil2007}. The most striking
confirmation of the electronic structure proposed in\cite{santos-2007}
came from the observation, with scanning tunneling spectroscopy, of
two low energy Van-Hove peaks in the density of states, with a strongly
angle dependent energy difference; these were identified with the
occurrence of two saddle points in the band structure\cite{Li2010_short}. 

The continuum description was originally developed for a specific
family of commensurate structures, dense in the low angle limit, in
which the relative displacement of corresponding Dirac points in each
layer, $\Delta\mathbf{K}=\mathbf{K}^{\theta}-\mathbf{K}$, ($\mathbf{K}^{\theta}$
is obtained from $\mathbf{K}$ by a rotation of the twist angle between
the layers) is not a reciprocal lattice vector of the Moiré super-lattice;
as a consequence there is no direct hopping matrix element between
these two Dirac points. Mele\cite{PhysRevB.81.161405} considered
the commensurability conditions more generally, and pointed out the
existence of another family of structures in which $\Delta\mathbf{K}$
is reciprocal lattice vector of the Moiré super-lattice. This matrix
element between the Dirac points of the two layers should then give
rise to a significant gap, raising the possibility of quite different
physics from the one discussed in \cite{santos-2007}.

Meanwhile, several authors \cite{laissardiere2010,PhysRevB.81.165105,PhysRevB.82.121407}
addressed the physics at very low twist angles ($\theta\lesssim1\text{º )}$,
finding significant deviations from some of the results presented
in our previous work. The continuum model is similar to a quasi-free
electron calculation, where the kinetic energy scale is $\hbar v_{F}\Delta K=2\hbar v_{F}K\sin(\theta/2)$
and the periodic potential scale in given by the inter-layer hopping.
The original calculation included a minimum set of plane waves, an
approximation which in only valid if the kinetic energy scale dominates.

In this work we review and extend the continuum model to address these
issues. We are able to present a complete analytical calculation of
all the Fourier components of the spatially modulated hopping for
any family of commensurate structures in the low angle limit. The
structures considered by Mele turn out to be quasi-periodic repetition
of simpler structures of the type we originally considered. The Fourier
components of the hopping amplitude that could lead to a gap, vanish
as the angle decreases, due to an interference effect, whereas other
amplitudes saturate, essentially ensuring that the low angle physics
of all commensurate structures in the one we discussed previously. 

The complete characterization of the Fourier components of the interlayer
hopping amplitude allows us to extend the treatment of the continuum
model to very small angles. The Fermi velocity vanishes at an angle
$\theta\sim1\text{º}$ in very good agreement with the results obtained
from band structure calculations\cite{laissardiere2010,PhysRevB.81.165105,PhysRevB.82.121407};
an almost dispersioneless band appears at this angle, corresponding
to localized states around regions of $AA$ stacking \cite{laissardiere2010}.
Using the continuum model, with only the dominant Fourier amplitude,
Bistritzer and MacDonald\cite{ISI:000293129900015} showed that at
even smaller angles the Fermi velocity becomes non-zero again, vanishing
at a series of {}``magic angles'', of which $\theta\sim1\text{º}$
is the first in the series. We present a simple explanation of this
observation based on the differences of the band structures of pure
$AB$ and pure $AA$ stacked bilayers. 

In section~\ref{sec:Geometry-main} we review the geometry of commensurate
structures in the twisted bilayer in order to establish notation and
present a new derivation of the results obtained by Mele\cite{PhysRevB.81.161405}
and Shallcross \emph{et. al.}\cite{PhysRevB.81.165105}. We formulate
the continuum model in section~\ref{sec:The-Continuum-Model} and
present an analytical formulation of the calculation of the Fourier
components of the spatially modulated inter-layer hopping, valid for
small angles and any kind of structure. The main results of the model
are presented in section~\ref{sec:The-continuum-model-results},
followed by a brief summary.

\section{Geometry of Commensurate Structures\label{sec:Geometry-main}}

The conditions for the commensurability of a Moiré pattern of two
rotated honeycomb lattices have already been considered by Mele\cite{PhysRevB.81.161405}
and Shallcross \emph{et. al.}\cite{PhysRevB.81.165105}. We review
this question, both to establish notation and to present an elementary
approach to this question, more directly based on the symmetries of
the hexagonal lattice. In this section we sketch the main argument,
leaving details for Appendix~\ref{sec:Geometry}.

The honeycomb (HC) lattice of graphene has an underlying Bravais lattice
with basis vectors which we choose as (lattice parameter $a=2.46\,\mathrm{\lyxmathsym{\AA}}$)\begin{subequations}\label{eq:basis}
\begin{eqnarray}
\mathbf{a}_{1} & = & a\left\{ \frac{1}{2},\frac{\sqrt{3}}{2}\right\} \label{eq:a1}\\
\mathbf{a}_{2} & = & a\left\{ -\frac{1}{2},\frac{\sqrt{3}}{2}\right\} \label{eq:a2}
\end{eqnarray}
\end{subequations}This lattice is made up of two sub-lattices, $A$
and $B$, where $A$ atoms occupy Bravais lattice nodes, and the $B$
are shifted by $\boldsymbol{\delta}_{1}=(\mathbf{a}_{1}+\mathbf{a}_{2})/3$:\begin{subequations}\label{AB_lattice_pos}
\begin{eqnarray}
\mathbf{r}_{A}(m,n) & = & m\mathbf{a}_{1}+n\mathbf{a}_{2}\label{eq:ra1}\\
\mathbf{r}_{B}(m,n) & = & \mathbf{r}_{A}(m,n)+\boldsymbol{\delta}_{1}\qquad m,n\in\mathbb{Z}.\label{eq:rb1}
\end{eqnarray}
\end{subequations}

In an $AB$ stacked bilayer there are two such lattices, vertically
displaced by $c=3.35\,\textrm{\AA},$ with the $B$ atoms of layer
2 ($B_{2}$) with the same horizontal positions as the $A$ atoms
in layer 1 ($A_{1})$, $\mathbf{r}_{B2}(m,n)=\mathbf{r}_{A1}(m,n)$.

In a twisted bilayer, the layers are rotated relative to each other.
We will assume we rotate layer 2 by an angle $\theta$, about a common
$A_{1}B_{2}$ horizontal position, that we take to be the origin.
A commensurate structure will occur if such a stacking $A_{1}B_{2}$
occurs elsewhere, say at $\mathbf{T}_{1}$; the rotation might as
well have been made about that second point, so $\mathbf{T}_{1}$
is a super-lattice translation, though not necessarily a primitive
vector. For $A_{1}B_{2}$ stacking to occur, a $B_{2}$ site must
rotate to a $A_{1}$ site,\emph{
\begin{equation}
k\mathbf{a}_{1}+l\mathbf{a}_{2}\to m\mathbf{a}_{1}+n\mathbf{a}_{2}\qquad k,l,m,n\in\mathbb{Z},\label{eq:ka1_la2}
\end{equation}
}which can only occur if 
\begin{equation}
k^{2}+l^{2}+kl=m^{2}+n^{2}+mn\label{eq:d_kl_d_mn}
\end{equation}
since $\left|k\mathbf{a}_{1}+l\mathbf{a}_{2}\right|^{2}=k^{2}+l^{2}+kl$. 

\begin{figure}
\subfloat[\label{fig:mag_geom1-1}]{\includegraphics[width=0.4\columnwidth]{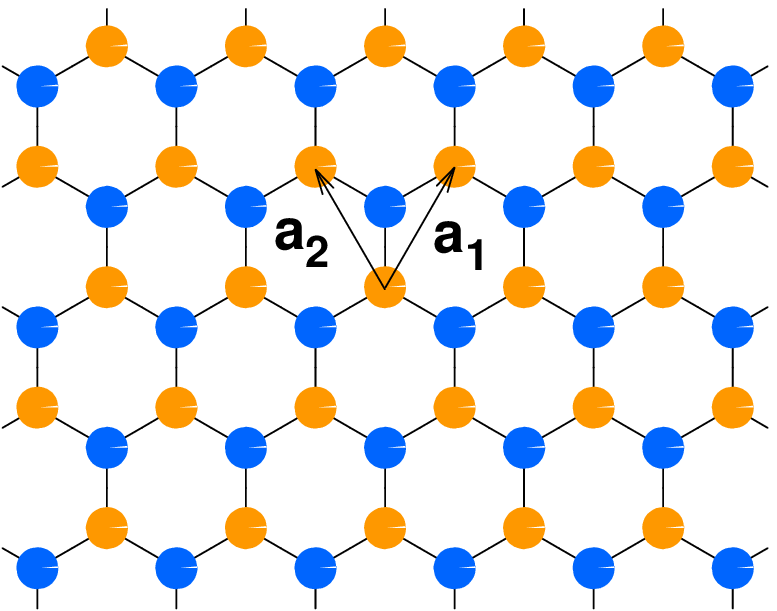}

}\subfloat[\label{fig:A-shell-12}]{\includegraphics[width=0.6\columnwidth]{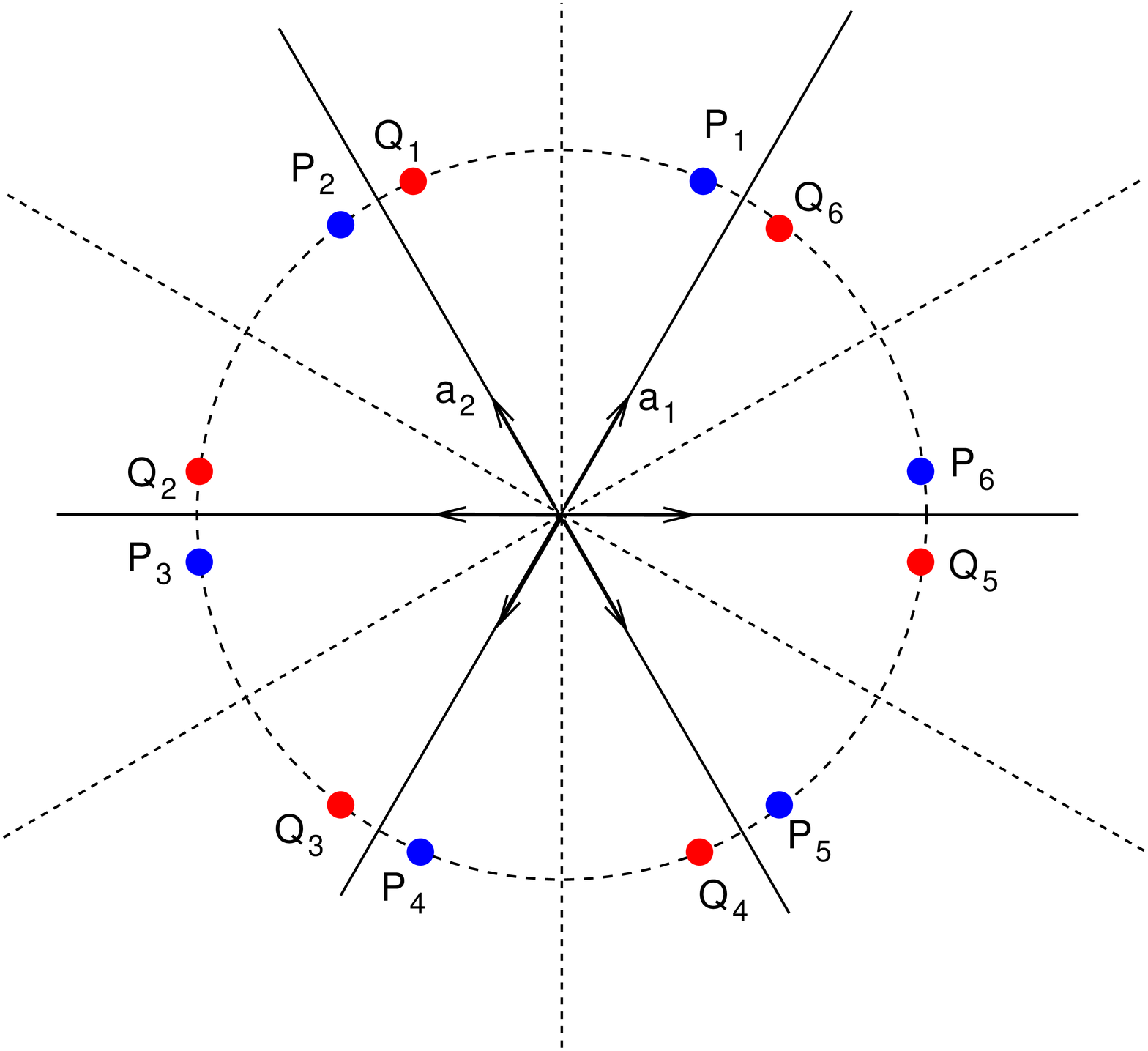}}

\caption{(a) Geometry of the Honeycomb lattice. (b) A shell of twelve Bravais
lattice sites, their position related by the rotation and reflections
symmetries of the hexagonal lattice.}
\end{figure}
Shallcross \emph{et. al. }in \cite{PhysRevB.81.165105} present a
detailed discussion of the solutions of this Diophantine equation.
The same conclusions can be reached by exploring the point symmetries
of the hexagonal lattice, namely the existence of a six-fold rotation
axis and of six reflection axis (the lines along the basis vectors
$\mathbf{a}_{1},$$\mathbf{a}_{2}$ and $\mathbf{a_{2}-}\mathbf{a}_{1}$,
and three axis at angles of $\pi/6$ with these). These symmetries
imply that a shell of Bravais lattice sites at a given distance from
the origin, must be built of groups of two sets of 6 sites, with position
vectors, $\mathbf{P}_{i}$ and $\mathbf{Q}{}_{i}$, $i=1,\dots,6$,
such as displayed in Fig.~(\ref{fig:A-shell-12}): the $\mathbf{P}_{i}(\mathbf{Q}_{i})$
lie at directions making an angle of $\pi/3$, and the two sets are
related to each other by reflection on the symmetry axis; these two
sets may degenerate into one if it occurs \emph{on the symmetry axes.
}Naturally, a rotation of layer 2 by the angle $\theta$ that brings
$\mathbf{P}_{1}\to\mathbf{Q}_{1}$ will leave six $A_{1}B_{2}$ sites
at the $\mathbf{Q}$ sites, each defining a lattice translation $\mathbf{T}_{i}$
of a commensurate structure (from origin to $\mathbf{Q}_{i}$). The
same can be said of the conjugate rotation $\theta'=\pi/3-\theta$
that maps $\mathbf{Q}_{6}\to\mathbf{P}_{1}$, in which case the lattice
translations are defined by the $\mathbf{P}_{i}$. Now, there may
be, at a given shell, more than one of these groups of symmetry related
sites. A shell of say 24 atoms will have two such groups $\mathbf{P}_{i}$,
$\mathbf{Q}_{i}$ and $\mathbf{R}_{i}$, $\mathbf{S}_{i}$. A rotation
that, say, maps $\mathbf{R}_{i}\to\mathbf{Q}_{i}$ must map $\mathbf{S}_{i}\to\mathbf{P}_{i}$
by symmetry, leaving us with 12 $A_{1}B_{2}$ sites at the same distance
from the origin: these lattice translations \emph{cannot be primitive
translations, }since the Bravais super-lattice is hexagonal by symmetry,
and only has\emph{ six} nearest neighbors. Thus, in order to find
all angles of commensuration, and the corresponding primitive vectors,
we need only consider rotations that map $\{\mathbf{P}_{i}\}\to\{\mathbf{Q}_{i}\}$
or $\{\mathbf{Q}_{i}\}\to\{\mathbf{P}_{i}\}$, where each of these
sets of six points is obtained from the other by reflection about
the symmetry axes.

These observations, and some elementary manipulations (see Appendix~\ref{sec:Geometry})
are sufficient to establish the following results for the possible
commensurate structures.

\emph{Angles}: the following equation, with $m$ and $r$ co-prime
positive integers, defines all possible angles of commensurate structures
with $0<\theta<\pi/3$:

\begin{equation}
\cos\theta(m,r)=\frac{3m^{2}+3mr+r^{2}/2}{3m^{2}+3mr+r^{2}},\label{eq:cos_theta_mr}
\end{equation}
\emph{Primitive~vectors}: the primitive vectors of the super-lattice
for a commensurate structure of angle $\theta(m,r)$ are:

i. If $\gcd(r,3)=1$, 
\begin{eqnarray}
\left[\begin{array}{c}
\mathbf{t}_{1}\\
\mathbf{t}_{2}
\end{array}\right] & = & \left[\begin{array}{cc}
m & m+r\\
-(m+r) & 2m+r
\end{array}\right]\left[\begin{array}{c}
\mathbf{a}_{1}\\
\mathbf{a}_{2}
\end{array}\right];\label{eq:primitive_vec_typeI}
\end{eqnarray}

ii. If $\gcd(r,3)=3$,
\begin{equation}
\left[\begin{array}{c}
\mathbf{t}_{1}\\
\mathbf{t}_{2}
\end{array}\right]=\left[\begin{array}{cc}
m+\frac{r}{3} & \frac{r}{3}\\
-\frac{r}{3} & m+\frac{2r}{3}
\end{array}\right]\left[\begin{array}{c}
\mathbf{a}_{1}\\
\mathbf{a}_{2}
\end{array}\right].\label{eq:primitive_vecs_typeII-1}
\end{equation}
These two types of structures can be distinguished both in real and
in reciprocal space\cite{PhysRevB.81.161405}. Using the results of
Appendix~\ref{sec:Geometry}, it is straightforward to show that
in the first case, $\gcd(r,3)=1$, the vertexes of the real-space
Wigner-Seitz (WS) cell of the super-lattice alternate between $B_{1}A_{2}$
sites and hexagon centers; in the second case, each corner of the
WS cell is an hexagon center of one layer and an atom of the other.
In the reciprocal space, the shift in the Dirac point of the rotated
layer, $\mathbf{K}^{\theta}-\mathbf{K}$, is a reciprocal lattice
vector only in the second case. Mele \cite{PhysRevB.81.161405}, who
first called attention to these two types of commensurate structures
refers to them as sub-lattice exchange even (SE-even) when $\gcd(r,3)=3$
and SE-odd when $\gcd(r,3)=1$.

\section{The Continuum Model\label{sec:The-Continuum-Model}}

The continuum description of the twisted bilayer was introduced by
the authors\cite{santos-2007} in 2007. A single graphene layer admits
an effective description in terms of the Dirac-Weyl equation for states
close to one of the Dirac points\cite{RevModPhys.81.109,PGN06b}.
We use this description for the intra-layer Hamiltonians in the twisted
bilayer, taking into account that layer 2 is rotated with respect
to layer~1 by $\theta$. We consider states near the Dirac point
$\mathbf{K}=4\pi(1,0)/3$ in layer~1 and $\mathbf{K}^{\theta}=(4\pi/3)(\cos\theta,\sin\theta)$
in layer~2. We denote by $\Psi_{i}(r)$, $i=1,2$ the two component
Dirac fields for each of the layers $i=1,2$, and write the momentum
as $\mathbf{K}+\mathbf{k}$ in layer~1 and $\mathbf{K}^{\theta}+\mathbf{k}$
in layer 2. 

In momentum space the intra-layer Hamiltonians are \cite{santos-2007}
\begin{eqnarray}
\mathcal{H}_{1} & = & \hbar\sum_{k}\Psi_{1,\mathbf{k}}^{\dagger}v_{F}\mathbf{\mbox{\ensuremath{\boldsymbol{\tau}}}\cdot k}\Psi_{1,\mathbf{k}}\label{eq:ham1_cont}\\
\mathcal{H}_{2} & = & \hbar\sum_{k}\Psi_{2,\mathbf{k}}{}^{\dagger}v_{F}\mathbf{\mbox{\ensuremath{\boldsymbol{\tau}}}^{\theta}\cdot}\mathbf{k}\Psi_{2,\mathbf{k}};\label{eq:ham2_cont}
\end{eqnarray}
the coordinate axes have been chosen to coincide with those of layer
1, $\mbox{\ensuremath{\boldsymbol{\tau}}}=(\tau_{x},\tau_{y}),\mbox{\ensuremath{\boldsymbol{\tau}}}^{\theta}=e^{+i\theta\tau_{z}/2}\mbox{\ensuremath{\boldsymbol{\tau}}}e^{-i\theta\tau_{z}/2}$,
and $\tau_{x}$ and $\tau_{y}$ are Pauli matrices. For the moment
we will ignore coupling between different Dirac valleys $\mathbf{K},\,\mathbf{K}^{\theta}$
and $\mathbf{K}'=\mathbf{-K},\,\mathbf{K}^{'\theta}=-\mathbf{K}^{\theta}$
; we will return to this point later.

To model the inter-layer coupling, $\mathcal{H}_{\perp}$, we retain
hopping from each site in layer 1 to the closest sites of layer 2
in either sub-lattice. We denote by $\boldsymbol{\delta}^{\beta'\alpha}(\mathbf{r})$
the horizontal (in-plane) displacement from an atom of layer 1, sub-lattice
$\alpha$($\alpha=A_{1},B_{1})$ and position $\mathbf{r}$, to the
closest atom in layer 2, sub-lattice $\beta'$ ($\beta'=A_{2},B_{'2}$).
The tight-binding inter-layer coupling is
\begin{equation}
\mathcal{H}_{\perp}=\sum_{i,\alpha,\beta'}t_{\perp}\left(\mathbf{\boldsymbol{\delta}}^{\beta'\alpha}(\mathbf{r}_{i})\right)c_{\alpha}^{\dagger}(\mathbf{r}_{i})c_{\beta'}\left(\mathbf{r}_{i}+\mathbf{\boldsymbol{\delta}}^{\beta'\alpha}(\mathbf{r}_{i})\right)+{\rm h.c.}\label{eq:h_perp1}
\end{equation}
where $t_{\perp}\left(\mathbf{\mathbf{\boldsymbol{\delta}}^{\alpha\beta}}(\mathbf{r})\right)\equiv t_{\perp}^{\alpha\beta}(\mathbf{r})$,
is the inter-layer, position dependent, hopping between $p_{z}$ orbitals
with a relative displacement $\mathbf{c}_{0}+\mathbf{\boldsymbol{\delta}}$,
and $c_{\alpha}(\mathbf{r})$ is the destruction operator for the
state in sub-lattice $\alpha$ at horizontal position $\mathbf{r}$.

Denoting by $\Delta\mathbf{K}=\mathbf{K}^{\theta}-\mathbf{K}$ the
relative shift between corresponding Dirac wave vectors in the two
layers, the usual replacement\cite{santos-2007} $c_{\alpha}(\mathbf{r})\to v_{c}^{1/2}\psi_{1,\alpha}(\mathbf{r})\exp(i\mathbf{K}\cdot\mathbf{r})$
leads to 
\begin{eqnarray}
\mathcal{H}_{\perp} & = & \sum_{\alpha\beta}\int d^{2}r\, t_{\perp}^{\beta\alpha}(\mathbf{r})e^{i\mathbf{K}^{\theta}\cdot\mathbf{\boldsymbol{\delta}}^{\beta\alpha}(r)}e^{i\Delta\mathbf{K}\cdot\mathbf{r}}\psi_{1,\alpha}^{\dagger}(\mathbf{r})\psi_{2,\beta}\left(\mathbf{r}\right)\nonumber \\
 & + & h.c..\label{eq:h_perp2}
\end{eqnarray}
We used $\psi_{\beta}(\mathbf{r}+\mathbf{\boldsymbol{\delta}}^{\beta\alpha}(\mathbf{r}))\approx\psi_{\alpha}(\mathbf{r})$
since the Dirac fields are slowly varying on the lattice scale.

In Fourier space it is convenient to define $\phi_{i,\mathbf{k},\alpha}$
as the Fourier component of $\psi_{i,\alpha}\left(\mathbf{r}\right)$
for momentum $\mathbf{k}\pm\Delta\mathbf{K}/2$, the plus sign applying
in layer 1 and the minus sign to layer 2. With this choice, the Dirac
fields $\phi_{i,\mathbf{k},\alpha}$ with the same $\mathbf{k}$ vector
in both layers correspond to the same plane waves in the original
lattice; the Dirac cones occur at $\mathbf{k}=-\Delta\mathbf{K}/2$
in layer 1 and $\Delta\mathbf{K}/2$ in layer 2. 

For commensurate structures, the function $t_{\perp}^{\alpha\beta}(\mathbf{r})\exp\left[i\mathbf{K}^{\theta}\cdot\mathbf{\boldsymbol{\delta}}^{\alpha\beta}(\mathbf{r})\right]$
is periodic and has nonzero Fourier components only at the vectors
$\mathbf{G}$ of the reciprocal lattice:
\begin{equation}
\tilde{t}_{\perp}^{\alpha\beta}(\mathbf{G})=\frac{1}{V_{c}}\int_{uc}d^{2}r\, t_{\perp}^{\alpha\beta}(\mathbf{r})e^{i\mathbf{K}^{\theta}\cdot\mathbf{\boldsymbol{\delta}}^{\alpha\beta}(r)}e^{-i\mathbf{G}\cdot\mathbf{r}}.\label{eq:t_perp_G}
\end{equation}
The integral is over the unit cell of the super-lattice, of area $V_{c}$.

With these definitions the low energy effective Hamiltonian, near
$\mathbf{K}$, is
\begin{eqnarray}
\mathcal{H} & = & \hbar\sum_{k,\alpha\beta}\phi_{1,\mathbf{k},\alpha}^{\dagger}v_{F}\mbox{\ensuremath{\boldsymbol{\tau}}}_{\alpha\beta}\cdot\left(\mathbf{k}+\frac{\Delta\mathbf{K}}{2}\right)\phi_{1,\mathbf{k},\beta}\nonumber \\
 & + & \hbar\sum_{k,\alpha,\beta}\phi_{2,\mathbf{k},\alpha}^{\dagger}v_{F}\mathbf{\mbox{\ensuremath{\boldsymbol{\tau}}}}_{\alpha\beta}^{\theta}\cdot\left(\mathbf{k}-\frac{\Delta\mathbf{K}}{2}\right)\phi_{2,\mathbf{k},\beta}\nonumber \\
 & + & \left(\sum_{\alpha,\beta}\sum_{\mathbf{k},\mathbf{G}}\tilde{t}_{\perp}^{\beta\alpha}(\mathbf{G})\phi_{1,\mathbf{k}+\mathbf{G},\alpha}^{\dagger}\phi_{2,\mathbf{k},\beta}+h.c.\right)\label{eq:complete_ham2}
\end{eqnarray}
Before proceeding it is perhaps worthwhile to remark that, including
other interlayer hopping amplitudes, does not alter this description
in a fundamental way. We would still arrive at a Hamiltonian similar
to the one of Eq.~(\ref{eq:h_perp1}), but the hopping $t_{\perp}^{\alpha\beta}(\mathbf{r})\exp\left[i\mathbf{K}^{\theta}\cdot\mathbf{\boldsymbol{\delta}}^{\alpha\beta}(\mathbf{r})\right]$
would be replaced by a more complicated expression. 

In this formulation, this problem is similar to that of a quasi-free
electron band problem, because each layer has been reduced to a continuum,
so that the only periodicity remaining in the problem is that of the
Moiré super-lattice. The most important parameters are then the Fourier
amplitudes, $\tilde{t}_{\perp}^{\beta\alpha}(\mathbf{G})$ defined
by Eq.~(\ref{eq:t_perp_G}).

The implications of Mele's discussion of SE-even structures\cite{PhysRevB.81.161405}
can now be clearly stated. In the SE-odd, $r=1$, structures we discussed
in 2007, $\Delta\mathbf{K}=(2\mathbf{G}_{1}+\mathbf{G}_{2})/3$ \emph{is}
\emph{not }a reciprocal lattice vector of the Moiré. There is no matrix
element coupling between the Dirac cones $\mathbf{K}$ and $\mathbf{K}^{\theta}$
of the two layers. There is, in fact, a matrix element coupling the
different valleys, since $\mathbf{K}^{'\theta}-\mathbf{K}$ is a reciprocal
lattice vector; but this wave-vector has magnitude $\mathcal{O}(1/a)$,
and for Moirés with large periods, $L\gg a$, $t_{\perp}^{\beta\alpha}(\mathbf{r})$
is very slowly varying on the graphene lattice scale, and one would
expect such matrix elements to be very small. But, as Mele pointed
out, for an SE-even structure, ($\gcd(r,3)=3$), $\Delta\mathbf{K}=r(\mathbf{G}_{1}+\mathbf{G}_{2})/3$
\emph{is }a reciprocal lattice vector of magnitude of order $\mathcal{O}(1/L)$
and there seems to be no \emph{à priori }reason to neglect it. It
lifts the degeneracy between the two Dirac points and leads to a gap.
A complete analysis of the Fourier amplitudes, to which we now turn,
will allow us to resolve this issue.

\section{Calculation of Fourier Amplitudes}

\subsection{Structures with $r=1.$\label{sub:Structures-with-r1}}

We begin by considering the calculation of Fourier amplitudes for
$r=1$ structures. Surprisingly, for small angles, the amplitudes
for other structures can be reduced to these. 

In Reference\cite{santos-2007} we stated that in the low angle limit,
and for an $r=1$ structure, the dominant amplitudes are given by
the results of Table~\ref{tab:fourier_amplitudes}.
\begin{table}
\begin{centering}
\begin{tabular}{|c||c|c|c|}
\hline 
$\mathbf{G}$ & 0 & $\mathbf{-G}_{1}$ & $-\mathbf{G}_{1}-\mathbf{G}_{2}$\tabularnewline
\hline 
\hline 
$\tilde{t}_{\perp}^{BA}(\mathbf{G})$ & $\tilde{t}_{\perp}$ & $\tilde{t}_{\perp}$ & $\tilde{t}_{\perp}$\tabularnewline
\hline 
$\tilde{t}_{\perp}^{AB}(\mathbf{G})$ & $\tilde{t}_{\perp}$ & $e^{-i2\pi/3}\tilde{t}_{\perp}$ & $e^{i2\pi/3}\tilde{t}_{\perp}$\tabularnewline
\hline 
$\tilde{t}_{\perp}^{AA}(\mathbf{G})$ & $\tilde{t}_{\perp}$ & $e^{i2\pi/3}\tilde{t}_{\perp}$ & ${e^{-i2\pi/3}\tilde{t}}_{\perp}$\tabularnewline
\hline 
$\tilde{t}_{\perp}^{BB}(\mathbf{G})$ & $\tilde{t}_{\perp}$ & ${e^{i2\pi/3}\tilde{t}}_{\perp}$ & $e^{-i2\pi/3}\tilde{t}_{\perp}$\tabularnewline
\hline 
\end{tabular}
\par\end{centering}

\caption{The first and second line express exact results. In the next two lines
these results have corrections of order $a/L$ where $L$ is the period
of the super lattice; $\tilde{t}_{\perp}$ is real.\label{tab:fourier_amplitudes}}
\end{table}
We now give a complete justification of this statement, and show how
one can calculate analytically \emph{all }amplitudes for low angles.
We begin by showing how certain symmetries imply relations between
the horizontal shifts $\delta^{BA}(\mathbf{r})$ for different sub-lattices.

As stated in Section~\ref{sec:Geometry-main}, three of the six vertexes
of the WS cell are $B_{1}A_{2}$ sites: for instance, 
\[
\mathbf{R}=\frac{2\mathbf{t}_{1}-\mathbf{t}_{2}}{3}=m\mathbf{a}_{1}+\boldsymbol{\delta}_{1}=(m+1)\mathbf{a}'_{1}-\boldsymbol{\delta}'_{1}.
\]
Since the origin is a $A_{1}B_{2}$ site, $\mathbf{R}$ is simultaneously
a $A_{1}\to B_{1}$ and a $B_{2}\to A_{2}$ translation. Therefore,
if there is an $A_{1}$ site at $\mathbf{r}$ and $B_{2}$ site at
$\mathbf{r}+\delta^{BA}(\mathbf{r})$, there will be to a $B_{1}$
site at $\mathbf{r}+\mathbf{R}$ and $A_{2}$ site at $\mathbf{r}+\mathbf{R}+\boldsymbol{\delta}^{BA}(\mathbf{r})$,
implying, 
\begin{eqnarray*}
\mathbf{r}+\boldsymbol{\delta}^{BA}(\mathbf{r})+\mathbf{R} & = & \mathbf{r}+\mathbf{R}+\boldsymbol{\delta}^{AB}(\mathbf{r}+\mathbf{R}),
\end{eqnarray*}
and $\boldsymbol{\delta}^{BA}(\mathbf{r})=\boldsymbol{\delta}^{AB}(\mathbf{r}+\mathbf{R}).$ 

A somewhat more involved symmetry of this structure, namely, invariance
under reflection about the origin, sub-lattice exchange ($A_{1}\leftrightarrow B_{1}$,
$A_{2}\leftrightarrow B_{2}$) and translation by\textbf{ $\mathbf{R}=m\mathbf{a}_{1}+\boldsymbol{\delta}_{1}$,}
leads to a similar relation $\boldsymbol{\delta}^{BB}(\mathbf{r})=-\boldsymbol{\delta}^{AA}(-\mathbf{r}+\mathbf{R})$.\textbf{
}These symmetries are exact and imply the following relations for
the Fourier amplitudes:\begin{subequations}\label{eq:tAA_BB_tBA_AB}
\begin{eqnarray}
\tilde{t}_{\perp}^{AB}(\mathbf{G}) & = & e^{-i\mathbf{G}\cdot\mathbf{R}}\tilde{t}_{\perp}^{BA}(\mathbf{G})\label{eq:t_AB_tBA}\\
\tilde{t}_{\perp}^{BB}(\mathbf{G}) & = & e^{-i\mathbf{G}\cdot\mathbf{R}}\left(\tilde{t}_{\perp}^{AA}(\mathbf{G})\right)^{*}.\label{eq:tBB_tAA}
\end{eqnarray}
\end{subequations}With \textbf{$\mathbf{G}=k\mathbf{G}_{1}+l\mathbf{G}_{2}$,
}we get $\mathbf{G}\cdot\mathbf{R}=2\pi(2k-l)/3$.

The WS cell also has three vertexes which are hexagon centers (see
Section~\ref{sec:Geometry-main}); one such vertex is $\left(\mathbf{t}_{1}+\mathbf{t}_{2}\right)/3$
for $r=1$ structures. This means that \textbf{$\mathbf{R}=\left(\mathbf{t}_{1}+\mathbf{t}_{2}\right)/3+\boldsymbol{\delta}_{1}$
}is a $A_{1}$ site and $\mathbf{R'}=\left(\mathbf{t}_{1}+\mathbf{t}_{2}\right)/3+\boldsymbol{\delta}'_{1}$
is an $A_{2}$ site, and so $\boldsymbol{\delta}_{AA}(\mathbf{R}+\boldsymbol{\delta}_{1})=\boldsymbol{\delta}'_{1}-\boldsymbol{\delta}{}_{1}\sim\mathcal{O}(\theta)$.
If this were exactly zero, $\mathbf{R}+\boldsymbol{\delta}_{1}$ would
be a $A_{1}\to A_{1}$ and $B_{2}\to A_{2}$ translations, implying
\[
\delta_{AA}(\mathbf{r})=\delta_{BA}(\mathbf{r}-\mathbf{R}-\boldsymbol{\delta}_{1})+\mathcal{O}(\theta).
\]
 This leads to 
\begin{equation}
\tilde{t}_{\perp}^{AA}(\mathbf{G})\approx e^{-i\mathbf{G}\cdot\left(\mathbf{R}+\boldsymbol{\delta}_{1}\right)}\tilde{t}_{\perp}^{BA}(\mathbf{G})\label{eq:tAA_tBA}
\end{equation}
As before, $\mathbf{G}=k\mathbf{G}_{1}+l\mathbf{G}_{2}$ and $\mathbf{G}\cdot\left(\mathbf{R}+\boldsymbol{\delta}_{1}\right)=2\pi(k+l)/3+\mathcal{O}(1/L)$.

These three relations, Eqs.(\ref{eq:tAA_BB_tBA_AB}) and (\ref{eq:tAA_tBA}),
express all amplitudes in terms of $\tilde{t}_{\perp}^{BA}(\mathbf{G})$
and have been thoroughly confirmed by numerical evaluation of the
Fourier amplitudes by calculating the integrals of Eq.~(\ref{eq:t_perp_G})
as a lattice sum. For the specific values of $\mathbf{G}$ considered
in Table~\ref{tab:fourier_amplitudes} they lead to the phase factors
relating amplitudes for different sub-lattices.

Let us now consider the expression for $\tilde{t}_{\perp}^{BA}(\mathbf{G})$,
and write it as a lattice sum, 
\begin{equation}
\tilde{t}_{\perp}^{BA}(\mathbf{G})=\frac{1}{N_{c}}\sum_{i\in uc}t_{\perp}\left[\boldsymbol{\delta}^{BA}(\mathbf{r}_{i})\right]e^{i\mathbf{K}^{\theta}\cdot\delta_{BA}(r_{i})}e^{-i\mathbf{G\cdot}\mathbf{r}_{i}}.\label{eq:t_BA_1}
\end{equation}
In terms of $\mathbf{G}':=\Delta\mathbf{K}+\mathbf{G}$, we get
\begin{eqnarray}
\tilde{t}_{\perp}^{BA}(\mathbf{G}) & = & \frac{1}{N_{c}}\sum_{i\in uc}t_{\perp}\left[\boldsymbol{\delta}^{BA}(\mathbf{r}_{i})\right]\nonumber \\
 & \times & e^{i\mathbf{K}^{\theta}\cdot(\mathbf{r}_{i}+\delta_{BA}(r_{i}))}e^{-i\mathbf{K}\cdot\mathbf{r}_{i}}e^{-i\mathbf{G}'\cdot\mathbf{r}_{i}}\label{eq:t_BA_2}
\end{eqnarray}
In the WS cell of an $r=1$ structure this simplifies because $\exp\left[i\mathbf{K}^{\theta}\cdot\left(\mathbf{r}+\boldsymbol{\delta}^{BA}(\mathbf{r}_{i})\right)\right]\exp\left[-i\mathbf{K}\cdot\mathbf{r}\right]=1$
for all sites. It turns out, and this is a property that is exclusive
to $r=1$ structures, that the $B_{2}$ site closest to $A_{1}$ at
$\mathbf{r}_{A}=m\mathbf{a}_{1}+n\mathbf{a}_{2}$ is at $\mathbf{r}_{B'}=m\mathbf{a}_{1}'+n\mathbf{a}_{2}'$
(same $m$ and $n$), so that $\mathbf{K}^{\theta}\cdot\mathbf{r}_{B'}=\mathbf{K}\cdot\mathbf{r}_{A}$.
As a result, 
\begin{equation}
\tilde{t}_{\perp}^{BA}(\mathbf{G})=\frac{1}{N_{c}}\sum_{i\in uc}t_{\perp}\left[\boldsymbol{\delta}^{BA}(\mathbf{r}_{i})\right]e^{-i\mathbf{G}'\cdot\mathbf{r}_{i}}\label{eq:tab_simple}
\end{equation}
Since $\mathbf{r}_{B_{2}}:=\mathbf{r}_{A_{1}}+\boldsymbol{\delta}^{BA}(\mathbf{r}_{A_{1}})=\mathbb{R}(\theta)\cdot\mathbf{r}_{A}$
, where $\mathbb{R}(\theta)$ is the rotation matrix, and for small
angles, $\mathbb{R}(\theta)=\mathbf{1}+d\boldsymbol{\omega}\times\mathbf{r}$,
we get $\left|\boldsymbol{\delta}_{BA}(\mathbf{r})\right|=\theta r.$
We can therefore approximate the sum of Eq.~\ref{eq:tab_simple}
as an integral
\begin{eqnarray*}
\tilde{t}_{\perp}^{BA}(\mathbf{G}) & = & \frac{1}{N_{c}}\sum_{i\in uc}t_{\perp}\left[\boldsymbol{\delta}^{BA}(\mathbf{r}_{i})\right]e^{-i\mathbf{G}'\cdot\mathbf{r}_{i}}\\
 & \approx & \frac{1}{N_{c}\sigma}\int_{uc}\!\! d^{2}r\, t_{\perp}(\theta r)e^{-iG'r\cos\phi}
\end{eqnarray*}
To simplify, we replace the hexagonal unit cell with a circle of the
same area $N_{c}\sigma=\sqrt{3}L^{2}/2$, where $L$ is the super-lattice
parameter. The radius of the circle is $R_{\mathrm{ws}}=\left(\sqrt{3}/2\pi\right)^{1/2}L$
and 
\begin{eqnarray}
\tilde{t}_{\perp}^{BA}(\mathbf{G}) & = & \frac{2}{\sqrt{3}L^{2}}\int_{0}^{R_{\mathrm{ws}}}\!\! dr\, rt_{\perp}\left(\frac{r}{L}\right)\int_{0}^{2\pi}d\phi e^{iG'r\cos\phi}\nonumber \\
 & = & \frac{4\pi}{\sqrt{3}}\int_{0}^{\left(\frac{\sqrt{3}}{2\pi}\right)^{1/2}}\!\! dx\, xt_{\perp}(x)J_{0}(G'Lx)\label{eq:analytical_tBA}
\end{eqnarray}
where $J_{0}(x)$ is a Bessel function.

To calculate this integral we need to parametrize the hopping between
$p_{z}$ orbitals as a function of the horizontal shift $\boldsymbol{\delta}.$
We express it in the Slater-Koster parameters, $V_{pp\sigma}(d)$
and $V_{pp\pi}(d)$, where $d$ is the distance between the two atomic
centers, $d=\sqrt{c_{0}^{2}+\delta^{2}}$. For the $d$ dependence
of $V_{pp\sigma}(d)$ and $V_{pp\pi}(d)$ we used the parametrization
of ref.~\cite{PhysRevB.53.979-b_short}; $V_{pp\pi}(a_{0}/\sqrt{3})$,
is the in-plane nearest neighbor hopping, $t$, and $V_{pp\sigma}(c_{0})$
if the inter-layer hopping, $t_{\perp}$, in an $AB$ stacked bilayer.
The contribution of $V_{pp\pi}$ turns out to be negligible, and $t_{\perp}(\delta)$
is proportional to $t_{\perp}$: for $\delta=a_{0}/\sqrt{3}$, the
carbon-carbon distance in a layer, $t_{\perp}(\delta)/t_{\perp}\approx0.4$. 

\begin{figure}
\noindent \begin{centering}
\includegraphics[width=1\columnwidth]{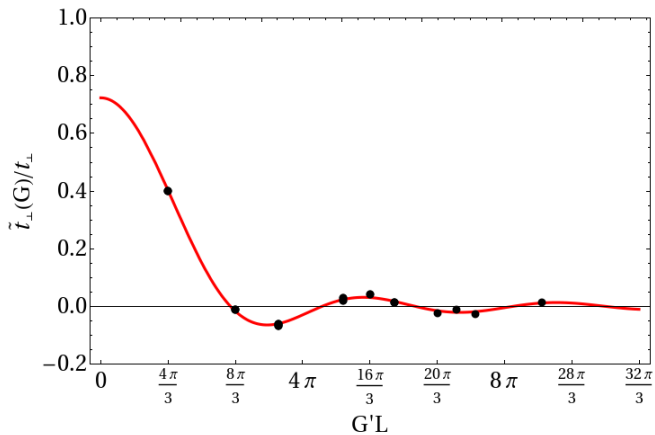}
\par\end{centering}

\caption{The $\tilde{t}_{\perp}^{BA}(\mathbf{G})/t_{\perp}$ as a function
of $G'L$: the dots are numerically calculated values for a $(m,r)=(10,1)$
structure, with $\theta=3.15\text{º}$, and the red line is the integral
of Eq.~(\ref{eq:analytical_tBA}). \label{fig:analytical_TBA}}
\end{figure}

With this parametrization, we represent the amplitude as a function
of $G'L$ in Fig.~\ref{fig:analytical_TBA}. If $t_{\perp}(\delta)$
were constant, the integral would be proportional to $J_{1}(G'L)/(G'L)$
and decay as $(G'L)^{-3/2}.$ This is actually the way this amplitude
decays, as could be seen by plotting $G'^{3/2}$ times the integral.
We have calculated numerically, as lattice sums, several amplitudes,
using Eq.~(\ref{eq:t_BA_1}); Fig.~\ref{fig:analytical_TBA} shows
that the analytical approximation to $\tilde{t}_{\perp}^{BA}(\mathbf{G})$
gives an excellent account of the values found numerically. 

These results are worthy of the following comments: 

(i) The three reciprocal lattice vectors selected in Table~\ref{tab:fourier_amplitudes},
$\mathbf{G}=0$,\textbf{ $\mathbf{G}=-\mathbf{G}_{1}$ }and $\mathbf{G}=-\mathbf{G}_{1}-\mathbf{G}_{2}$,
all have $G'L=4\pi/3$. The corresponding values of $\tilde{t}_{\perp}^{BA}(\mathbf{G})$
are equal, $\tilde{t}_{\perp}^{BA}(\mathbf{G})=0.4t_{\perp}$; all
other reciprocal lattice vectors have larger values of $G'$, and
the amplitudes are correspondingly smaller; these other amplitudes
were ignored in Refs.\cite{ISI:000293129900015,santos-2007}.

(ii) For a general $\mathbf{G}=k\mathbf{G}_{1}+l\mathbf{G}_{2},$
$\mathbf{G}'=(k+2/3)\mathbf{G}_{1}+(l+1/3)\mathbf{G}_{2}.$ Since
$\mathbf{G}_{i}\propto1/L$ , $G'L$ becomes independent of the angle
or rotation. The amplitudes for a given $(k,l)$ become independent
of angle for small angles,  tending to the values given by our analytical
approximation. 

(iii) This complete characterization of the Fourier Amplitudes, allows
one, in principle, to include in the calculation of the spectrum as
many plane waves as necessary to achieve convergence. The characteristic
energy from the in-plane motion is $\hbar v_{F}\Delta K\sim0.190\,\theta$,
with the energy in $\mathsf{eV}$ and the angle in degrees, and for
small angles one requires more plane waves than those used in Ref.\cite{santos-2007}.
The physics of these small angle structures has been widely discussed
recently in the literature and we will use these results to discuss
it in the framework of the continuum model. But, before that, we consider
the calculation of the Fourier amplitudes in other families of commensurate
structures.

\subsection{Importance of $r=1$ structures}

In this section we show that, in the small angle limit, the $r=1$
structures are special, and determine the physics of all types of
commensurate structures.

\begin{figure}
\subfloat[\label{fig:Lsin_theta_vs_theta-1}]{\includegraphics[width=1\columnwidth]{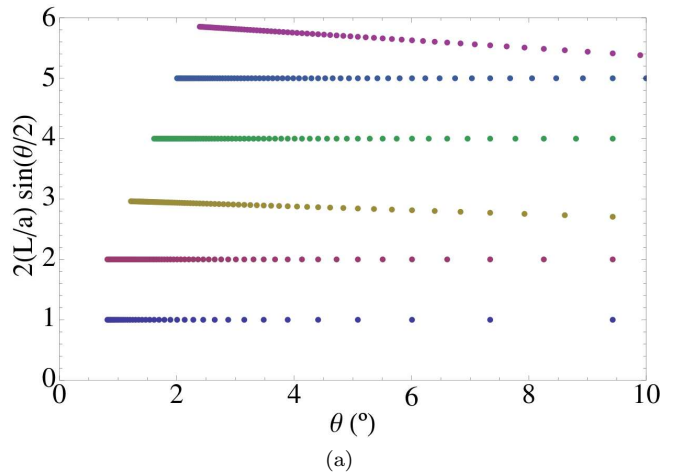}

}

\subfloat[\label{fig:type-II_type-I-1}]{\begin{centering}
\includegraphics[width=0.9\columnwidth]{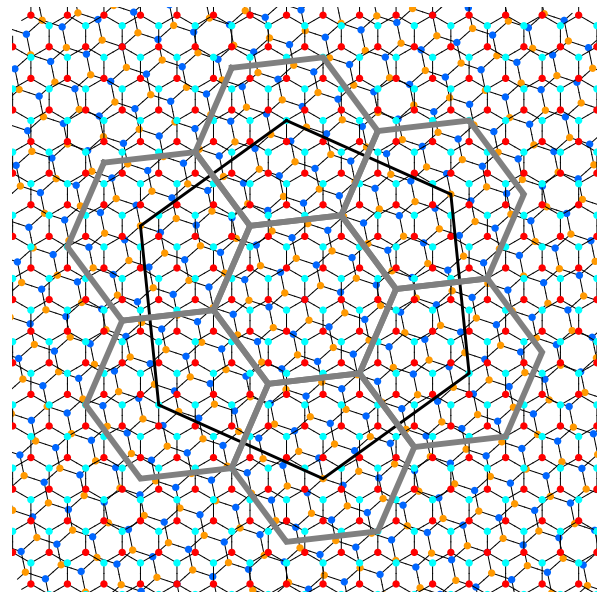}
\par\end{centering}

}

\caption{(a) $2(L/a)\sin(\theta/2)$ vs $\theta$. The various lines correspond
to different values of $r$; the lower line corresponds to the structures
with $r=1$. (b) A SE-even structure is almost periodic with the period
of a structure with $r=1$; here is a $(m,r)=(7,3)$ is shown overlaid
with the Wigner-Seitz Cells of $(m,r)=(2,1);$ the black hexagon is
the true unit cell of the structure. }
\end{figure}

In STM images\cite{RK93}, Moiré patterns appear to satisfy the following
relation between period and angle of rotation: $L=a/\left[2\sin(\theta/2)\right]$.
For a general $(m,r)$ structure, \begin{subequations}\label{sin_theta_L}
\begin{eqnarray}
\sin\left(\frac{\theta(m,r)}{2}\right) & = & \frac{1}{2}\frac{r}{\sqrt{3m^{2}+3mr+r^{2}}}\label{eq:sin_theta}\\
L(m,r) & =a & \sqrt{3m^{2}+3mq+q^{2}},\label{eq:period_L}
\end{eqnarray}
\end{subequations}where $q=r/\gcd(r,3)$, so the above relation is
only satisfied for $r=1$. The plot $2L\sin(\theta/2)/a$ as a function
$\theta$, in Fig.~\ref{fig:Lsin_theta_vs_theta-1}, makes this clear.
Remark that all these families of super-lattices, with different values
of $r$, are dense as $\theta\to0.$ This means that a very small
change in $\theta$, with little effect in the structure in real space,
can nevertheless change $L$ by an arbitrary large factor. The implication
is that, for very small angles, all commensurate structures are \emph{almost
periodic repetitions }of structures with $r=1.$ That is seen very
clearly by inspecting visually a few Moiré patterns {[}see Fig.~(\ref{fig:type-II_type-I-1}){]}. 

Let us show this explicitly for a SE-even structure, $(m,r)$ with
$r=3r'$. At one of the corners of the Wigner-Seitz cell, 
\begin{equation}
\mathbf{r}:=\frac{\mathbf{t}_{1}+\mathbf{t}_{2}}{3}=m\boldsymbol{\delta}_{1}+\frac{r}{3}\mathbf{a}_{2}=m\boldsymbol{\delta'}_{1}+\frac{r}{3}\mathbf{a}'_{1};\label{eq:corner_ws}
\end{equation}
 If $m\mod3=1$, like in the $(7,3)$ structure in Fig.~\ref{fig:type-II_type-I-1},
this site has $B_{1}$ atom of layer 1 and a hexagon center of layer
2. Therefore, at $\mathbf{r}-\boldsymbol{\delta}_{1}$ there is an
$A_{1}$ site and at $\mathbf{r}-\mathbf{\boldsymbol{\delta}}'_{1}$,
a $B_{2}$ one. This implies that $\boldsymbol{\delta}^{BA}(\mathbf{r})=\boldsymbol{\delta}'_{1}-\boldsymbol{\delta}{}_{1}=\mathcal{O}(\theta)$.
If this were zero, $\mathbf{r}$ would be a lattice translation of
the Moiré. The corresponding structure would be of SE-odd with $m'=(m-1)/3$
and $r'=r/3$. In real space, a SE-even structure $(m,r)$, with $m-1$
divisible by 3, is then very similar to a SE-odd with $(m',r')=((m-1)/3,r/3)$.
In the following paragraphs we refer to these two lattices as $\mathcal{L}$
(SE-even) and $\widetilde{\mathcal{L}}$ (SE-odd).

\begin{figure}
\includegraphics[width=1\columnwidth]{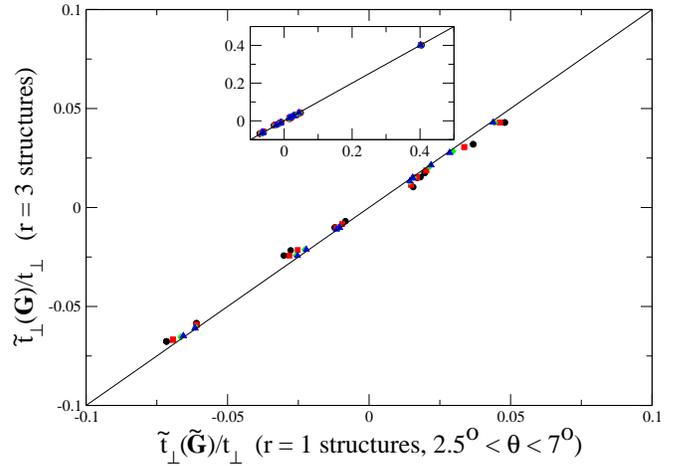}

\caption{Comparison of Fourier amplitudes of pairs of structures: each point
has an $x$ coordinate $\tilde{t}_{\perp}^{\beta\alpha}(\mathbf{G})$
for $\left((m-1/3,r/3\right)$ and a $y$ coordinate $\tilde{t}_{\perp}^{\beta\alpha}(\mathbf{G})$
for a SE-even structure $(m,r$). $\mathbf{\widetilde{\mathbf{G}}}=k\mathbf{\widetilde{\mathbf{G}}}_{1}+l\mathbf{\widetilde{\mathbf{G}}}_{2}$
and $\mathbf{G}=(2k-l)\mathbf{G}_{1}+(k+l)\mathbf{G}_{2}$. According
to Eq.~(\ref{eq:beautiful_result}), these amplitudes should be equal.
The line is $y=x$, not a fit. The inset has an expanded scale to
include the dominant amplitudes ($k,l)=\{(0,0),\,(-1,0),\,(-1,-1)\}.$
The angles are in the range $2.5\text{º}<\theta<7.3\text{º}$.\label{fig:Comparison-of-Fourier-amplitudes}}
\end{figure}

Let us now relate the reciprocal lattice primitive vectors of $\mathcal{L}$
and $\widetilde{\mathcal{L}}$ . Using the results of Appendix~\ref{sec:Geometry}
one arrives at 

\begin{eqnarray}
\left[\begin{array}{c}
\widetilde{\mathbf{G}}{}_{1}\\
\widetilde{\mathbf{G}}{}_{2}
\end{array}\right] & = & \left(\left[\begin{array}{cc}
2 & 1\\
-1 & 1
\end{array}\right]+\mathcal{O}(\theta)\right)\left[\begin{array}{c}
\mathbf{G}{}_{1}\\
\mathbf{G}{}_{2}
\end{array}\right]\label{eq: SE-even_SE-odd_reciprocal}
\end{eqnarray}
Ignore, for the moment, the $\mathcal{O}(\theta)$ corrections. These
equations tell us that the real space basis $\mathcal{L}$, $\mathbf{t}_{1},\mathbf{t}_{2}$,
are linear combinations with integer coefficients of the basis of
$\widetilde{\mathcal{L}}$. In the present case we have:
\begin{equation}
\left[\begin{array}{c}
\mathbf{t}{}_{1}\\
\mathbf{t}{}_{2}
\end{array}\right]\approx\left[\begin{array}{cc}
2 & 1\\
-1 & 1
\end{array}\right]\left[\begin{array}{c}
\widetilde{\mathbf{t}}{}_{1}\\
\widetilde{\mathbf{t}}{}_{2}
\end{array}\right]\label{eq:SE-even_SE_odd_direct}
\end{equation}
In the calculation of $\tilde{t}_{\perp}^{\beta\alpha}(\mathbf{G})$
for the lattice with primitive vectors \textbf{$\mathbf{t}_{1},$$\mathbf{t}_{2},$
}we can take into account that the $\boldsymbol{\delta}^{\beta\alpha}(\mathbf{r}_{i})$
are (approximately) periodic in $\widetilde{\mathbf{t}}{}_{1}$ and
$\widetilde{\mathbf{t}}{}_{2}$, and split the sum over $\mathbf{r}_{i}$
in the unit cell $\mathcal{L}$, into a sum over the unit cell of
$\widetilde{\mathcal{L}}$, $\mathbf{r}_{i}'$ and a sum over the
$n_{c}$ unit cells of $\widetilde{\mathcal{L}}$ contained in the
unit cell $\mathcal{L}$: 

\begin{align}
\tilde{t}_{\perp}^{\beta\alpha}(\mathbf{G})= & \frac{1}{n_{c}}\sum_{\mathbf{R}_{n}}e^{-i\mathbf{G\cdot}\widetilde{\mathbf{T}}{}_{n}}\nonumber \\
\times\frac{1}{\widetilde{N}_{c}}\sum_{\mathbf{r}_{i}'\in\widetilde{uc}} & t_{\perp}\left[\delta_{\beta\alpha}(\mathbf{r}'_{i}+\widetilde{\mathbf{T}}{}_{n})\right]e^{i\mathbf{K}^{\theta}\cdot\delta_{\beta\alpha}(\mathbf{r'}_{i}+\mathbf{T}'_{n})}e^{-i\mathbf{G\cdot}\mathbf{r'}_{i}}\nonumber \\
\approx & \frac{1}{n_{c}}\sum_{\widetilde{\mathbf{T}}{}_{n}}e^{-i\mathbf{G\cdot}\widetilde{\mathbf{T}}{}_{n}}\nonumber \\
\times\frac{1}{\widetilde{N}_{c}}\sum_{\mathbf{r}_{i}'\in\widetilde{uc}} & t_{\perp}\left[\delta_{\beta\alpha}(\mathbf{r}'_{i})\right]e^{i\mathbf{K}^{\theta}\cdot\delta_{\beta\alpha}(\mathbf{r'}_{i})}e^{-i\mathbf{G\cdot}\mathbf{r'}_{i}}.\label{eq:sep_structure_factor.}
\end{align}
We achieved a factorization of $\tilde{t}_{\perp}^{\beta\alpha}(\mathbf{G})$\begin{subequations}\label{factor_t}
\begin{align}
\tilde{t}_{\perp}^{\beta\alpha}(\mathbf{G}) & =S(\mathbf{G})\left[\frac{1}{\widetilde{N}_{c}}\sum_{i\in\widetilde{uc}}t_{\perp}\left[\delta_{\beta\alpha}(\mathbf{r}_{i})\right]e^{i\mathbf{K}^{\theta}\cdot\delta_{\beta\alpha}(\mathbf{r}_{i})}e^{-i\mathbf{G\cdot}\mathbf{r}_{i}}\right]\label{eq:factor_t_1}\\
S(\mathbf{G}) & =\frac{1}{n_{c}}\sum_{\widetilde{\mathbf{T}}{}_{n}}e^{-i\mathbf{G\cdot}\widetilde{\mathbf{T}}{}_{n}}\label{eq:factor_t_2}
\end{align}
 \end{subequations}The second factor is $\tilde{t}_{\perp}^{\beta\alpha}(\mathbf{G})$
for the lattice $\widetilde{\mathcal{L}}$. As for the structure factor,
$S(\mathbf{G})$, note that, by definition, $\mathbf{G}\cdot\mathbf{T}=2m\pi$,
if $\mathbf{T}$ is a translation vector of $\mathcal{L}$ (periodic
boundary conditions), and, if $\mathbf{G=}\widetilde{\mathbf{G}}$,
a reciprocal vector of $\widetilde{\mathcal{L}}$, $\exp[i\mathbf{G}\cdot\widetilde{\mathbf{T}}{}_{n}]=1$.
Therefore we obtain, in this approximation,
\begin{equation}
\tilde{t}_{\perp}^{\beta\alpha}(\mathbf{G})=\left[\frac{1}{\widetilde{N}_{c}}\sum_{i\in\widetilde{uc}}t_{\perp}\left[\delta_{\beta\alpha}(\mathbf{r}_{i})\right]e^{i\mathbf{K}^{\theta}\cdot\delta_{\beta\alpha}(\mathbf{r}_{i})}e^{-i\mathbf{G\cdot}\mathbf{r}_{i}}\right]\delta_{\mathbf{G},\widetilde{\mathbf{G}}}\label{eq:beautiful_result}
\end{equation}
where $\widetilde{\mathbf{G}}$ is any reciprocal vector $\widetilde{\mathcal{L}}$.
This is a very important result:
\begin{description}
\item [{\textmd{(i)}}] it expresses the Fourier amplitudes of SE-even in
terms of those of structures with $r=1,$ which we calculated in section~\ref{sub:Structures-with-r1}; 
\item [{\textmd{(ii)}}] it states an approximate selection rule, that becomes
more accurate as the angle of rotation decreases, allowing us to identify
Fourier amplitudes that must tend to zero for small angles. 
\end{description}
We have checked this result by numerical calculation of $\tilde{t}_{\perp}^{\beta\alpha}(\mathbf{G})$
for various lattices using Eq.~(\ref{eq:t_BA_1}). In Fig.~\ref{fig:Comparison-of-Fourier-amplitudes},
each point on the plot has an $x$ coordinate equal to $\tilde{t}_{\perp}^{\beta\alpha}(\widetilde{\mathbf{G}})$,
$\widetilde{\mathbf{G}}=k\mathbf{\widetilde{\mathbf{G}}}{}_{1}+l\widetilde{\mathbf{G}}{}_{2}$,
for the $(m',r')=((m-1)/3,r/3)$ lattice, and a $y$ coordinate $\tilde{t}_{\perp}^{\beta\alpha}(\mathbf{G})$,
$\mathbf{G}=(2k-l)\mathbf{G}{}_{1}+(k+l)\mathbf{G}{}_{2}$ of the
SE-even $(m,r)$ lattice; Eq.~(\ref{eq:beautiful_result}) predicts
that these amplitudes should be equal and the agreement is excellent.

\begin{figure}
\includegraphics[width=1\columnwidth]{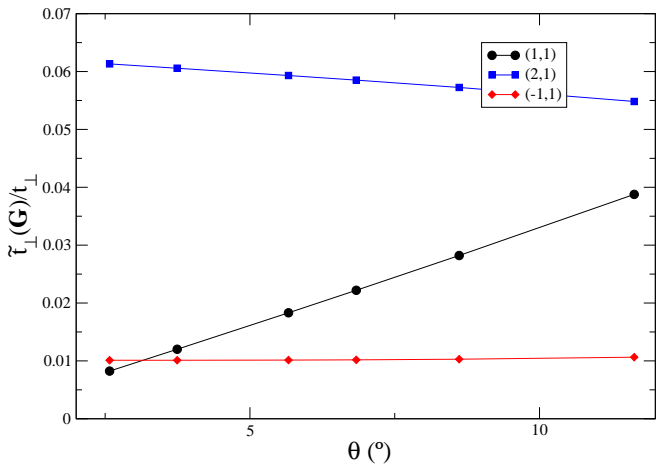}

\raggedright{}\caption{$\left|\tilde{t}_{\perp}^{\beta\alpha}(\mathbf{G})\right|/t_{\perp}$
for $\mathbf{G}=k\mathbf{G}_{1}+l\mathbf{G}_{2}$ for SE-even lattices,
with different angles of rotation. For $(k,l)=(-1,1)$ or $(2,1)$,
$\mathbf{G}$ is approximately equal to a reciprocal lattice vector
of a $r=1$ lattice and, $\tilde{t}_{\perp}^{\beta\alpha}(\mathbf{G})/t_{\perp}$is
almost constant; but for $(k,l)=(1,1)$, which corresponds to $\mathbf{G}=\Delta\mathbf{K}$,
the amplitude vanishes linearly with $\theta$.\label{fig:amplitude_1_1}}
\end{figure}

The second implication of Eq.~(\ref{eq:beautiful_result}) concerns
the behavior of Fourier Amplitudes for which $\mathbf{G}$ is not
a reciprocal lattice vector of $\widetilde{\mathcal{L}}$, the $r=1$
super-lattice. Of particular interest is $\mathbf{G}=\Delta\mathbf{K}=\mathbf{G}_{1}+\mathbf{G}_{2}$,
because it determines the magnitude of the gap in a SE-even structure.
In Fig.~\ref{fig:amplitude_1_1} we show that $\left|\tilde{t}_{\perp}^{\beta\alpha}(\Delta\mathbf{K})\right|\to0$
as $\theta\to0,$ as a result of the vanishing of the structure factor
$S(\mathbf{G})$. In other words, there is a destructive interference
in the sum of Eq.~(\ref{eq:t_BA_1}), because of the quasi-periodicity
of the hopping amplitudes inside the unit cell of the larger period
lattice. When $\mathbf{G}$ matches a reciprocal lattice vector $\widetilde{\mathbf{G}}$
of the smaller period lattice $S(\mathbf{G})\approx1$ (constructive
interference) the amplitudes saturate to finite values as $\theta\to0$. 

With this knowledge of the Fourier amplitudes for any structure, we
finally address the calculation of the low energy bands of a small
angle bilayer with a twist.

\section{The continuum Model at low angles\label{sec:The-continuum-model-results}}

In the absence of the inter-layer coupling, $\mathcal{H}_{\perp}$,
states with energy close to zero occur at $\mathbf{k}=-\Delta\mathbf{K}/2$
in layer~1 and $\mathbf{k}=+\Delta\mathbf{K}/2$ in layer~2. The
interlayer Hamiltonian $\mathcal{H}_{\perp}$ couples the states of
momentum $\mathbf{k}$ in layer~1 to states $\mathbf{k}-\mathbf{G}$,
in layer 2 with a matrix element $\tilde{t}_{\perp}^{\beta\alpha}(\mathbf{G})$.
The most important Fourier amplitudes, (of modulus $\tilde{t}_{\perp}=0.4t_{\perp}$),
in $r=1$ structures, occur for $\mathbf{G}=0$, $\mathbf{G}=-\mathbf{G}_{1}$,
and $\mathbf{G}=-\mathbf{G}{}_{1}-\mathbf{G}_{2}$ for which $G'L=4\pi/3$,
where $\mathbf{G}'=\mathbf{G}+\Delta\mathbf{K}$ {[}see Fig.~(\ref{fig:analytical_TBA}){]}.
Neglecting other Fourier amplitudes\cite{santos-2007}, the states
of momentum $\mathbf{k}$ in layer 1 are coupled directly only to
states of layer 2 of momentum $\mathbf{k}$, $\mathbf{k}+\mathbf{G}_{1}$
and $\mathbf{k}+\mathbf{G}_{1}+\mathbf{G}_{2}$; conversely the states
of momentum $\mathbf{k}$ in layer 2 only couple to states $\mathbf{k}$,
$\mathbf{k}-\mathbf{G}_{1}$ and $\mathbf{k}-\mathbf{G}_{1}-\mathbf{G}_{2}$.
To investigate the spectrum at a momentum $\mathbf{k}$ close to zero
energy, one can truncate the Hamiltonian to include only these six
momentum values (three for each layer) giving a $12\times12$ matrix
to diagonalize (3 momentum values, 2 layers and 2 sub-lattices)\cite{santos-2007}.
When $\mathbf{k}$ is close to the Dirac cone of one layer, the three
momentum values that it couples to lie at the same distance $\Delta K$
from the Dirac point of the opposing layer; we have zero energy states
coupling to two triplets of states at $\pm v_{F}\Delta K$. The spectrum
obtained from the diagonalization of the Hamiltonian matrix can be
interpreted in a perturbative way when $\tilde{t}_{\perp}/v_{F}\Delta K\ll1$.
This analysis was presented in previous works and will not be repeated
here\cite{santos-2007,Li2010_short}. The main conclusions were: (i)
the persistence of The Dirac cones, with linear dispersion; (ii) a
renormalization of the Fermi velocity, relative to the single layer,
which, in perturbation theory, was predicted as $\tilde{v}_{F}/v_{F}=1-9\left(\tilde{t}_{\perp}/(\hbar v_{F}\Delta K)\right)^{2}$($v_{F}$
is the single layer value); (iii) the appearance of two low energy
Van-Hove peaks due to the appearance of saddle points in the low energy
bands, arising from the mixing of the two Dirac cones.

In SE-even structures, however, there is a direct matrix element coupling
the two Dirac cones; will the physics change relative to SE-odd structures
due to the appearance of a gap? 

According to Eq.~(\ref{eq: SE-even_SE-odd_reciprocal}), the dominant
Fourier amplitudes, in this case, occur for $\mathbf{G}=0,$ $\mathbf{G}=-2\mathbf{G}_{1}-\mathbf{G}_{2}$
and $\mathbf{G}=-\mathbf{G}_{1}-2\mathbf{G}_{2}$, since these correspond
to $\widetilde{\mathbf{G}}=0,$-$\widetilde{\mathbf{G}}_{1},-\widetilde{\mathbf{G}}_{1}-\widetilde{\mathbf{G}}_{2}$;
on the other hand, $\Delta\mathbf{K}=r(\mathbf{G}_{1}+\mathbf{G}_{2})/3=\mathbf{G}_{1}+\mathbf{G}_{2}$.
Therefore, these three dominant amplitudes couple the Dirac point
of layer~1 to states of the layer~2 which are shifted from its Dirac
point by $-\Delta\mathbf{K}=-(\mathbf{G}_{1}+\mathbf{G}_{2})$ ,$-\Delta\mathbf{K}+2\mathbf{G}_{1}+\mathbf{G}_{2}=\mathbf{G}_{1}$
and $-\Delta\mathbf{K}+\mathbf{G}_{1}+2\mathbf{G}_{2}=\mathbf{G}_{2}$;
since the angle between $\mathbf{G}_{1}$ and $\mathbf{G}_{2}$ is
$2\pi/3$, these are three vectors of the same modulus, $\left|\Delta\mathbf{K}\right|$,
at $2\pi/3$ angles, and we recognize exactly the same situation as
discussed above for the $r=1$ structures: the degeneracy points of
each layer couple to two triplets at energies $\pm v_{F}\Delta K$.
It is true that, for this structure, there is a direct matrix element
coupling the two degeneracy points, corresponding to $\mathbf{G}_{1}+\mathbf{G}_{2}$,
which will lift the degeneracy and lead to a gap. However, as we saw
in Fig.~(\ref{fig:amplitude_1_1}), this matrix element decreases
with angle, and below 5º is under $5$~meV. 

One can say that the differences between the various types of structures
in momentum space are somewhat of a red herring. Two structures which
are almost identical in real space \emph{must }display similar physics.
The momentum space description can \emph{look }very different, but
the magnitudes of the Fourier amplitudes must ensure similar results. 

The perturbation theory in $\tilde{t}_{\perp}/(\hbar v_{F}\Delta K)$
clearly breaks down for very small angles, since, as we have seen
the numerator becomes constant of order $0.4t_{\perp}\sim0.1\,\mathsf{eV}$
and the denominator is $\hbar v_{F}\Delta K\sim0.190\times\theta\,\mathsf{eV}$
(angle in degrees). This has led some authors\cite{laissardiere2010,PhysRevB.81.165105,PhysRevB.82.121407}
to question the validity of the continuum description in the small
angle limit. Band structure calculations, while confirming the prediction
~of depressed Fermi velocity, $\tilde{v}_{F}/v_{F}=1-9\left(\tilde{t}_{\perp}/(\hbar v_{F}\Delta K)\right)^{2}$,
find deviations from it below about $\theta\approx5\text{º}.$ 

One should not however confuse the perturbative result with the continuum
model. In fact, the continuum model should be better at smaller angles,
since the scale of variation of the inter-layer hopping becomes larger.
What one must do, however, is to include a larger set of plane waves
in order to achieve convergence of the low energy spectrum. 

In the following, we present some results for small angles, obtained
by diagonalizing numerically the Hamiltonian of Eq.(\ref{eq:complete_ham2}),
truncated to a finite basis (largest matrix used of $168\times168$),
and including all the required Fourier amplitudes, as given by the
analytical expression of Eq.(\ref{eq:analytical_tBA}). This limit
has already been addressed by Bistritzer and MacDonald\cite{ISI:000293129900015}
in an approximation that includes only the dominant Fourier amplitudes.
Some of our calculations, particularly those of the density of states,
apparently require larger matrices for convergence than the ones that
these authors claimed to have used. 

\begin{figure}
\includegraphics[width=1\columnwidth]{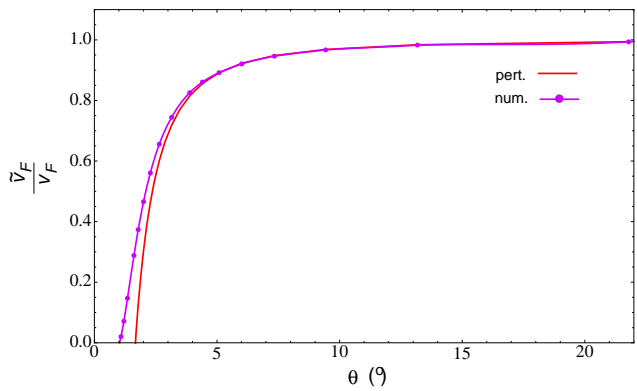}

\caption{Velocity renormalization, by perturbation theory in $\tilde{t}_{\perp}/(\hbar v_{F}\Delta K)$,
and by the continuum model with numerical diagonalization of the secular
equation, with a sufficient set of momentum values for convergence.
The latter calculation deviates from the perturbative results as $(\hbar v_{F}\Delta K)$
becomes comparable to $\tilde{t}_{\perp}$, but is in quite good agreement
with band structure calculations\cite{PhysRevB.82.121407,PhysRevB.81.165105,laissardiere2010}.\label{fig:Velocity-renormalization}}

\end{figure}

The results for the ratio of the Fermi velocity to the single layer
value, as function of $\theta$ are shown in Fig.~(\ref{fig:Velocity-renormalization}).
As expected, for small angles they deviate from the perturbative result,
and compare very well with the values obtained from band structure
calculations\cite{PhysRevB.82.121407,PhysRevB.81.165105,laissardiere2010}:
the Fermi velocity becomes zero at about $\theta\approx1\text{º}.$ 

\begin{figure}
\begin{centering}
\subfloat[\label{fig:contour-plot-m18}]{\begin{centering}
\includegraphics[width=0.48\columnwidth]{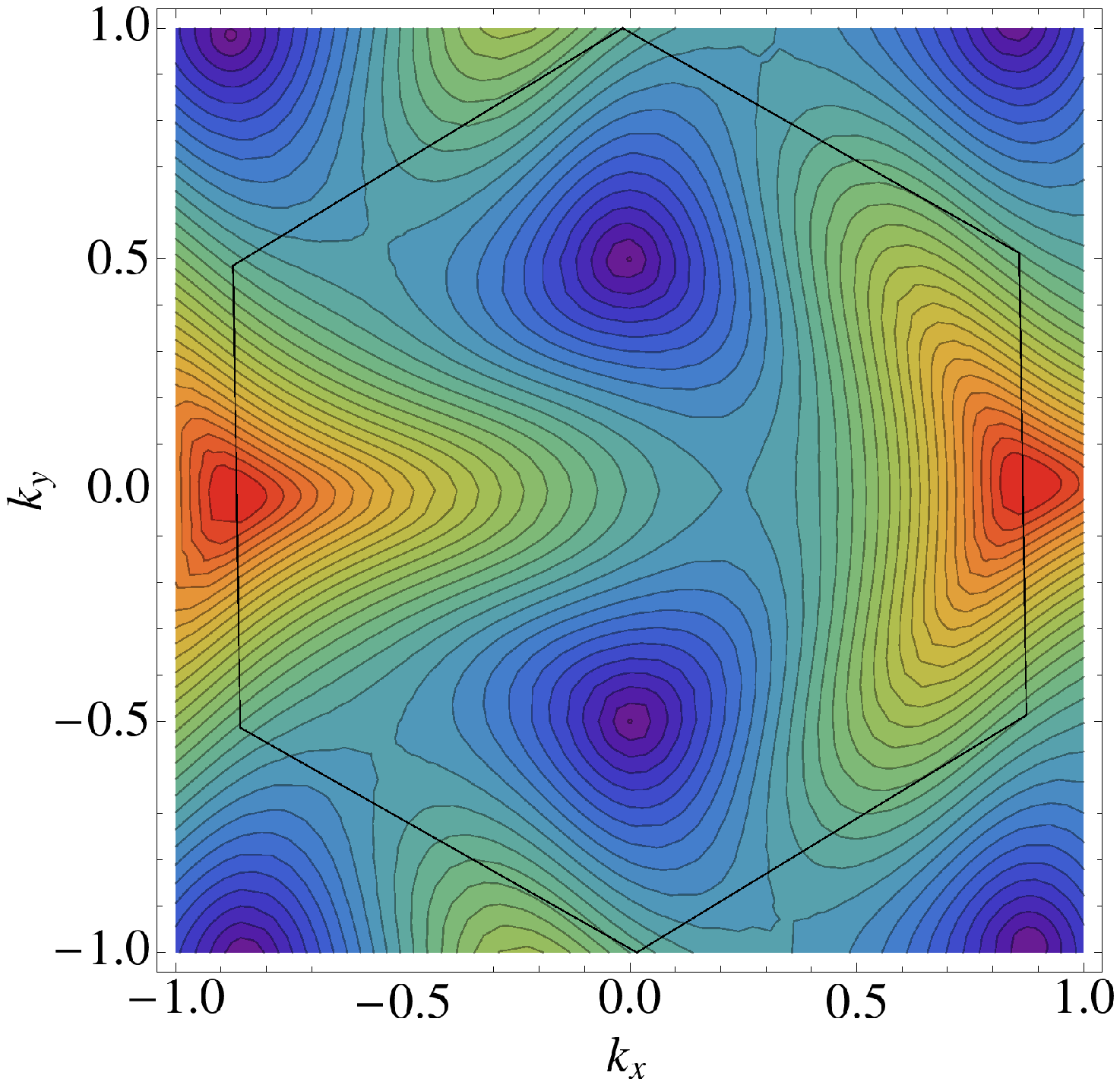}
\par\end{centering}

}\subfloat[\label{fig:contour-plot-m27}]{\begin{centering}
\includegraphics[width=0.5\columnwidth]{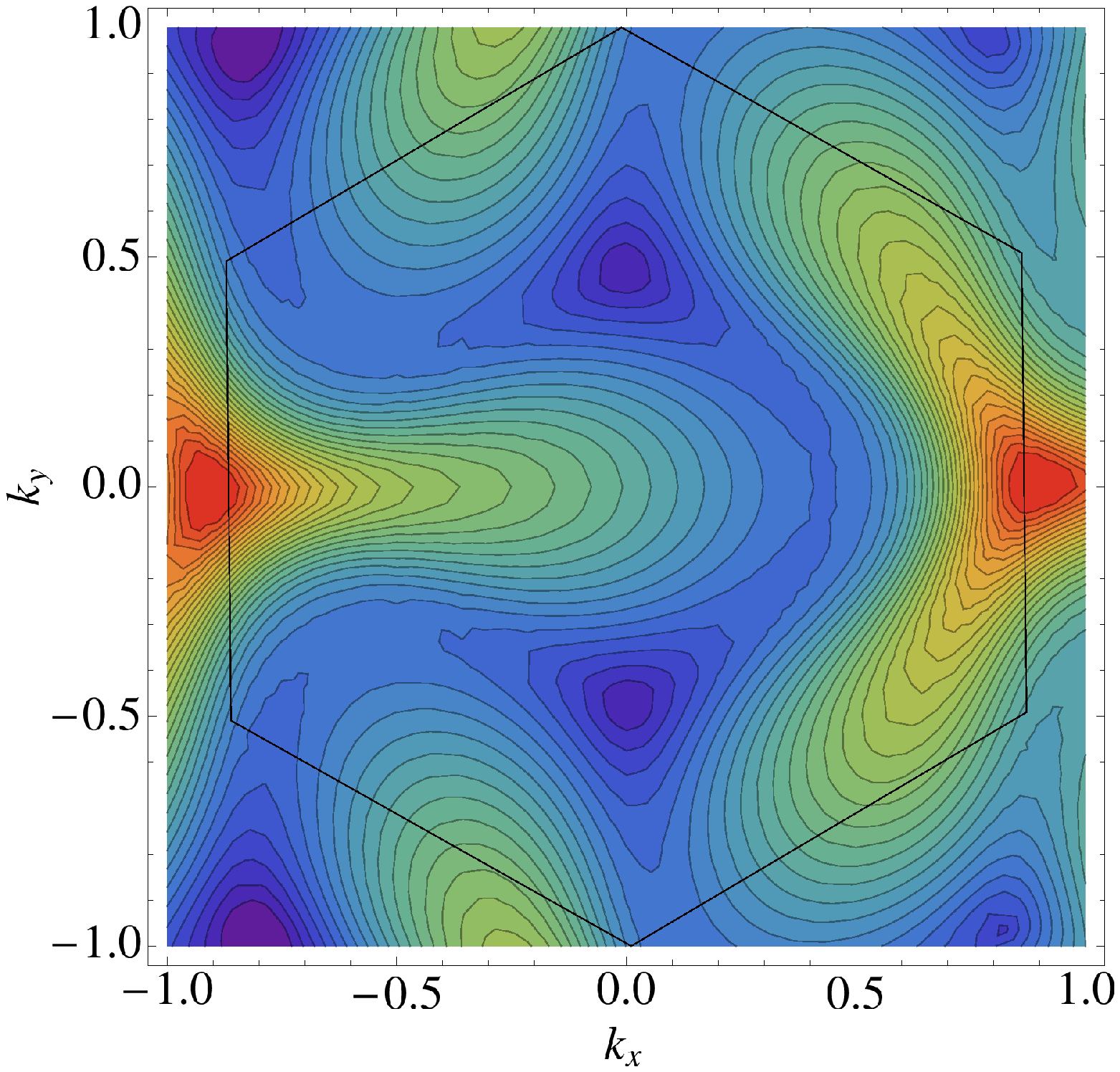}
\par\end{centering}

}
\par\end{centering}

\caption{Contour plot of the first positive energy band; the hexagon is the
First Brillouin Zone. (a) $\theta=1.79\text{º}$, the cones are visible,
but the saddle point is not located on the line joining the two Dirac
cones; (b) $\theta=1.20\text{º},$ the arc joining the cones through
the saddle point has become a very flat valley, and the cones are
no longer well defined.}
\end{figure}

In Fig.~\ref{fig:contour-plot-m18} we show a density plot of lowest
positive energy bands for $\theta=1.79\text{º}$ ($\widetilde{v}_{F}/v_{F}\approx0.3$);
the Dirac cones, as well as the saddle point between them, are clearly
visible; at even smaller angles, $\theta=1.20\text{º }$, the corresponding
plot shows an almost flat region in the arc joining the two Dirac
cones through the saddle point (Fig.~\ref{fig:contour-plot-m27});
the range of energies with linear dispersion becomes very small. 

\begin{figure}
\subfloat[\label{fig:DOSm18}]{\includegraphics[width=0.9\columnwidth]{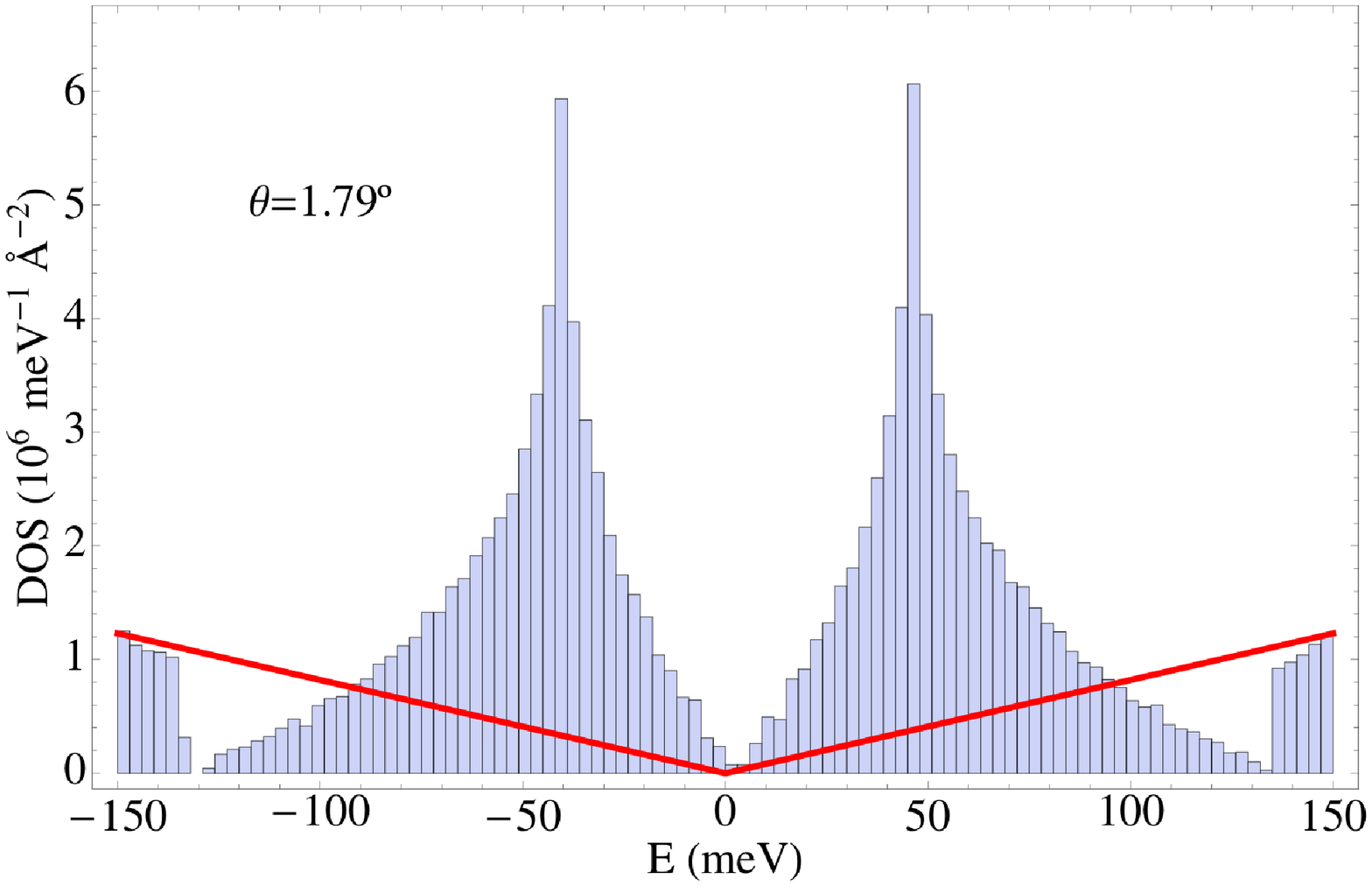}

}

\subfloat[\label{fig:dos_m27}]{\includegraphics[width=0.9\columnwidth]{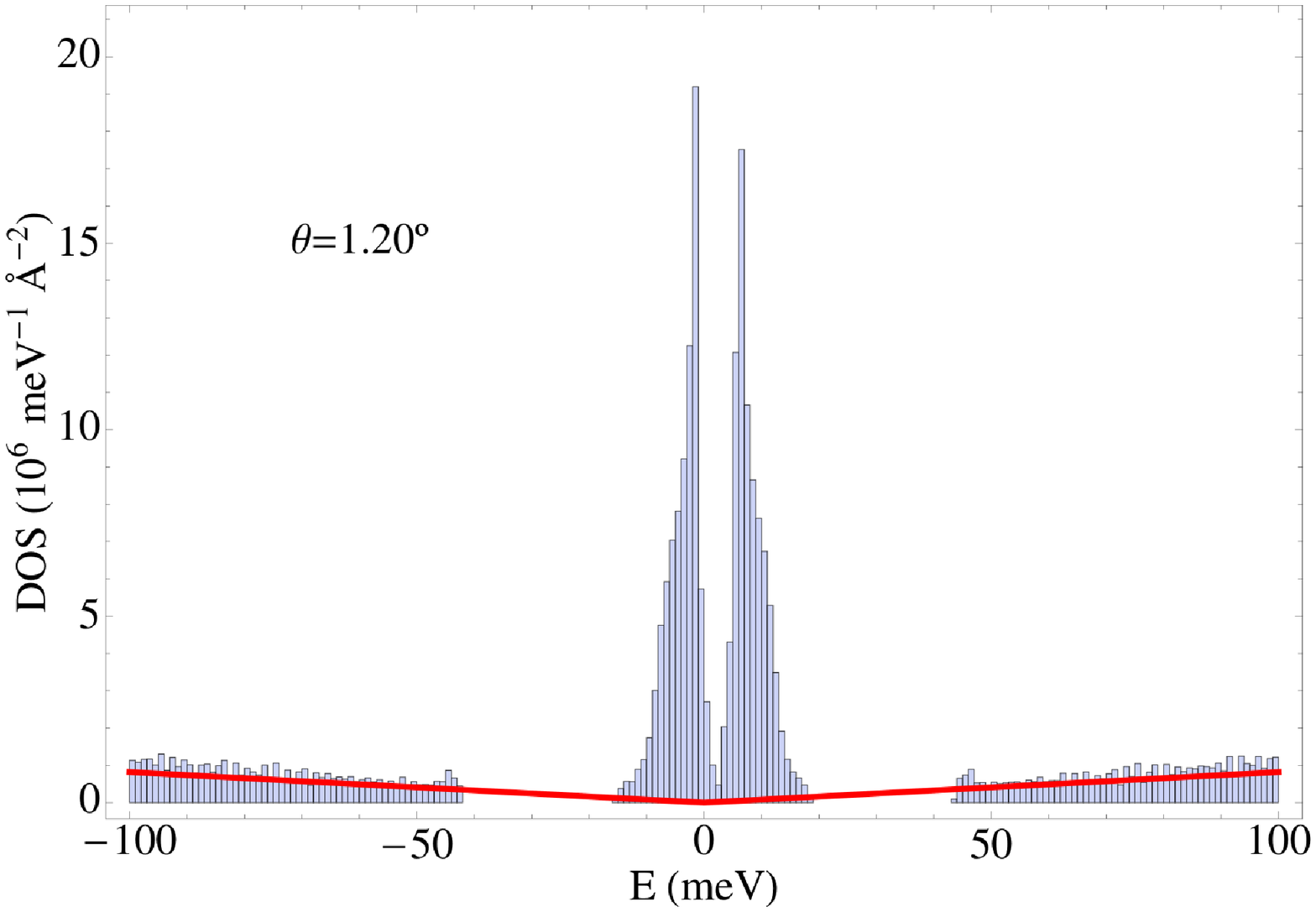}

}

\caption{Densities of states (DOS) for two angles; a) for $\theta=1.79\text{º }$the
cones are still well defined and the dispersion show the usual linear
dependence near zero energy; b) for $\theta=1.2\text{º }$there is
a finite density of states near zero energy and one cannot define
a Fermi velocity: the dispersion is no longer linear. The red line
is the DOS for two uncoupled layers.}
\end{figure}
The density of states (DOS) is a very convenient tool to check for
presence of Dirac cones in the band-structure. If the cones are present,
the DOS shows a linear dependence near zero energy, as can be clearly
seen in Fig.~\ref{fig:DOSm18} for $\theta=1.79\text{º};$ for $\theta=1.2\text{º}$,
one can still define a (very small) Fermi velocity, but one should
bear in mind that that the range of energies of linear dispersion
is contracted to a few \textsf{$\mathsf{meV}$}. 

\begin{figure}
\subfloat[\label{fig:DOSm30}]{\includegraphics[width=0.9\columnwidth]{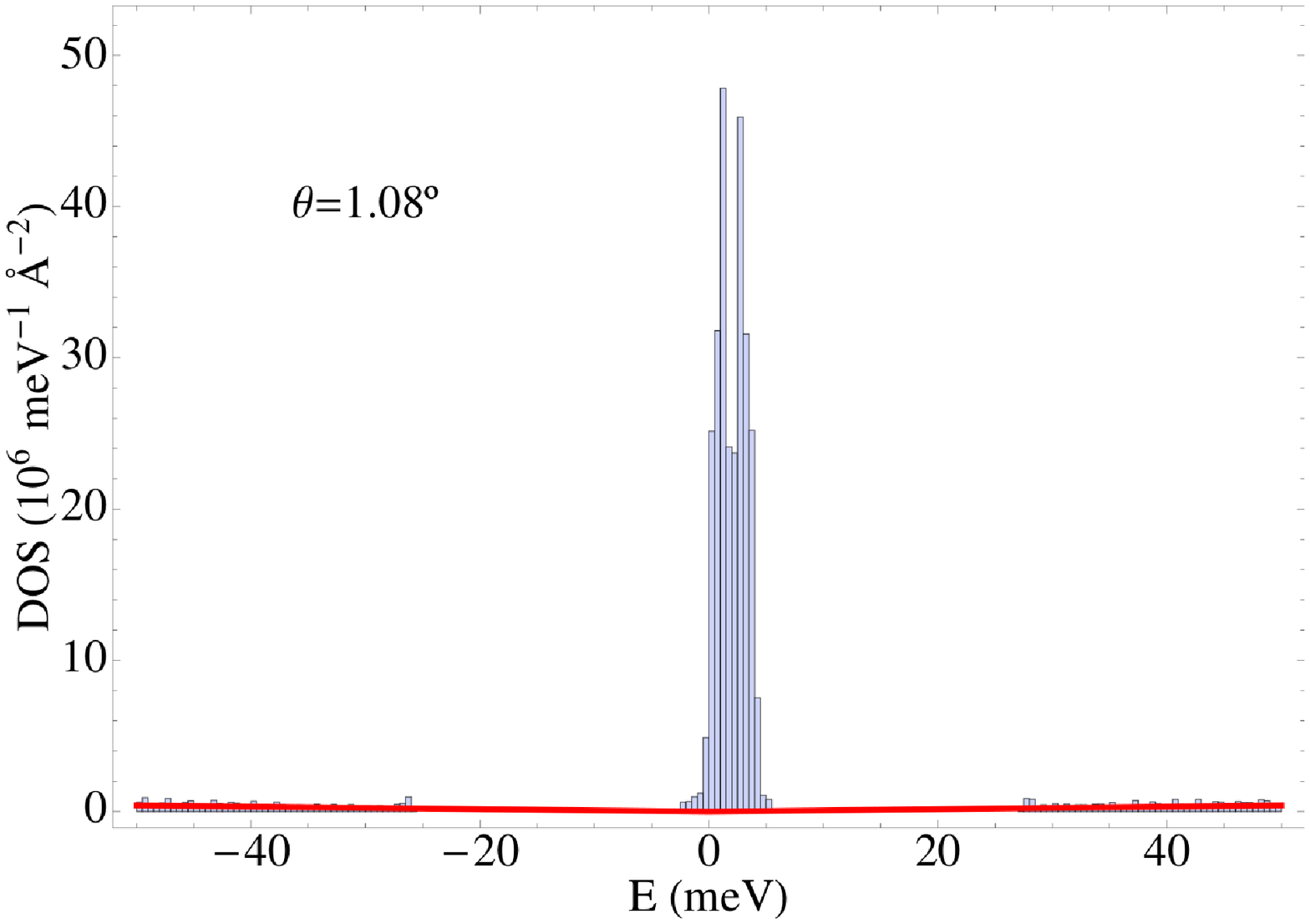}

}

\subfloat[\label{fig:dos_m40}]{\includegraphics[width=0.9\columnwidth]{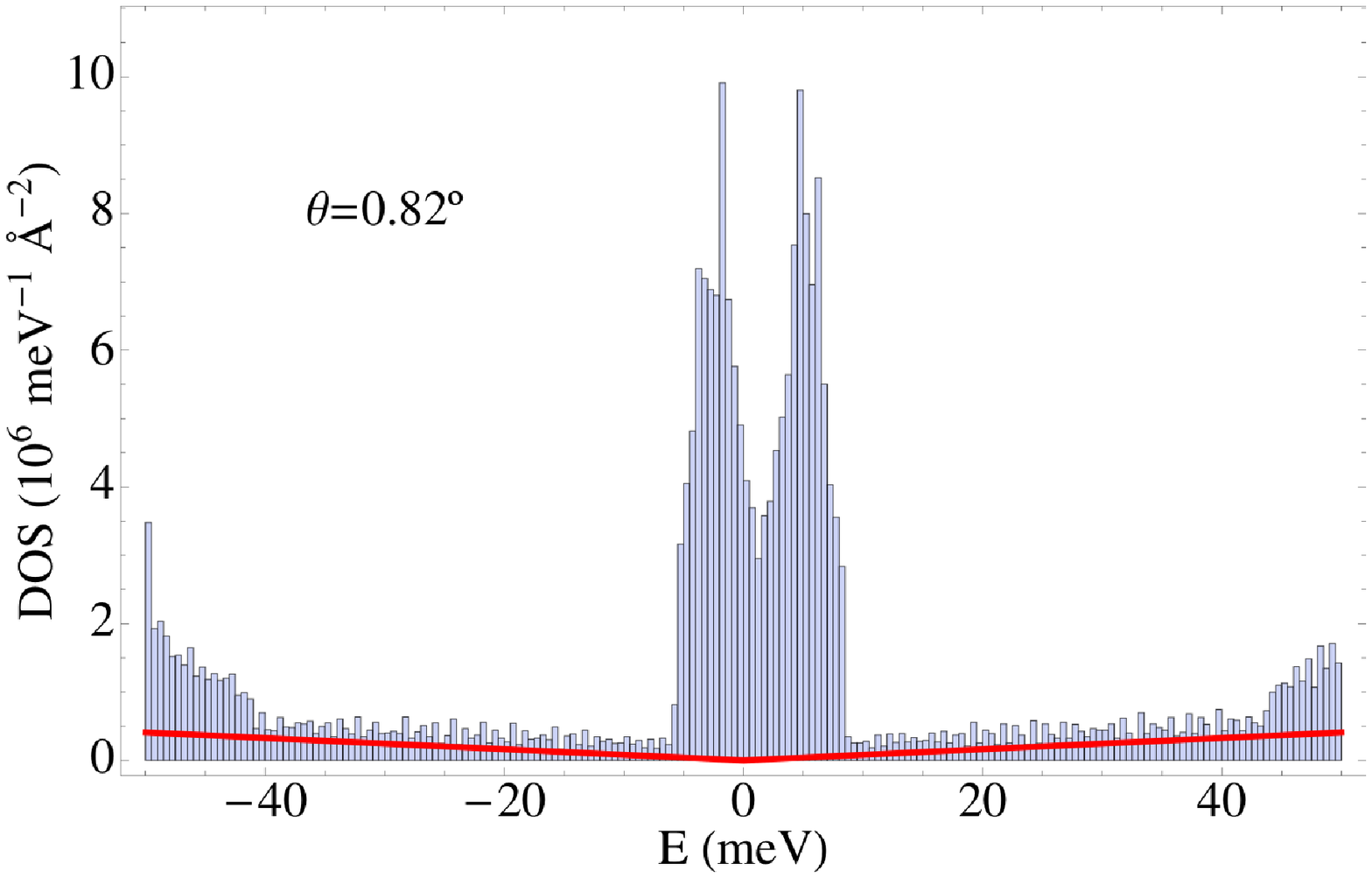}

}

\caption{Densities of states (DOS) for two small angles; a) for $\theta=1.08\text{º }$there
is a very sharp central peak, with a barely resolved two peak structure,
corresponding to a flat band of states localized in $AA$ stacking
regions of the unit cell. b) for $\theta=0.87\text{º }$ the central
band is broader, and still displays the two peak structure, although,
at the $\mathsf{meV}$ resolution there is a finite DOS between the
peaks, precluding the unambiguous definition of a Fermi velocity.\label{fig:DOS_smaller_angles}}
\end{figure}
 At an even smaller angle, $\theta=1.08\text{º}$, one observes a
sharp peak in the DOS at low energy, corresponding to an almost dispersioneless
band (Fig.~\ref{fig:DOSm30}), with a barely resolved two peak structure.
Surprisingly, if the angle decreases further, the central band broadens
(FIG:~\ref{fig:dos_m40}). This curious behavior was first found
by Bistritzer and MacDonald \cite{ISI:000293129900015} and characterized
as an oscillation of the Fermi velocity. In fact, at the $\mathsf{meV}$
resolution of the figure, the DOS is finite between the peaks. It
is not clear that a region of linear dispersion even exists, but,
if it does, it is so narrow, that we prefer to concentrate on this
curious variation of the width of the central peak. 
\begin{figure}
\subfloat[\label{fig:localDos_m18}]{\includegraphics[width=0.5\columnwidth]{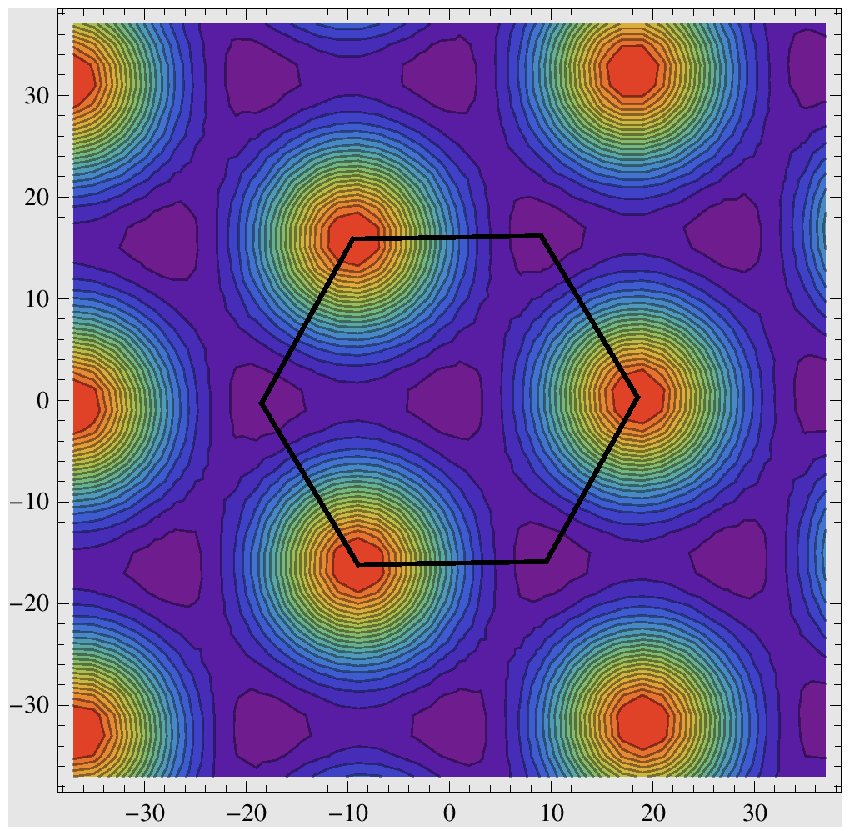}

}\subfloat[\label{fig:localdos_m_27}]{\includegraphics[width=0.5\columnwidth]{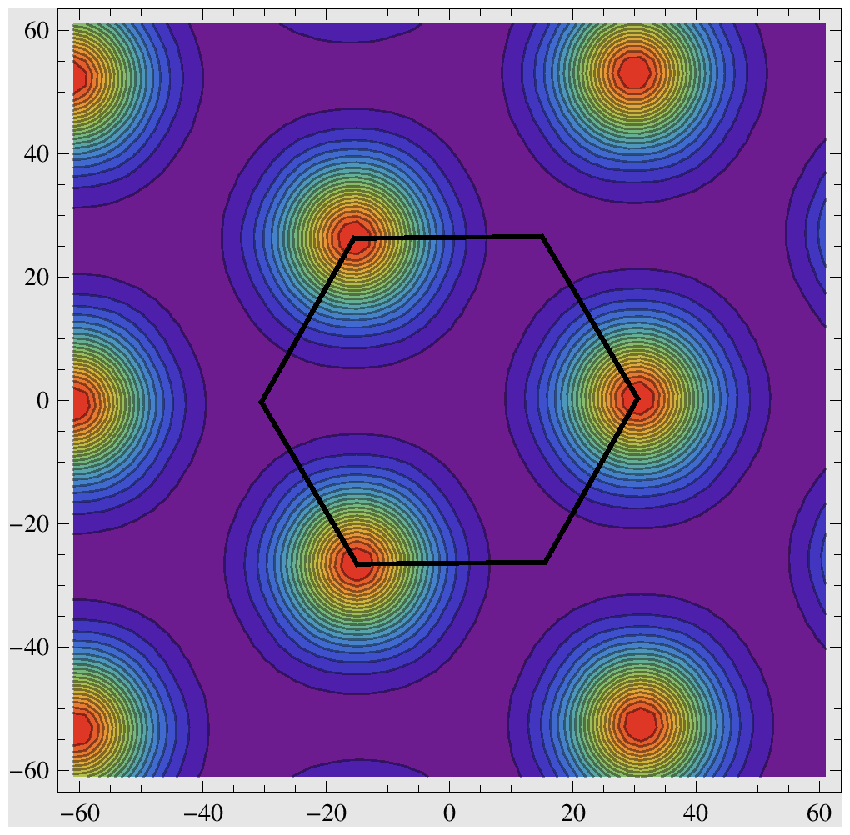}}

\caption{Density plots of the Local Density of states, in the Wigner-Seitz
Cell, integrated for $\left|\epsilon\right|<20\,\mathsf{meV}$: (a)
$\theta=1.78\text{º};$(b) $\theta=1.08\text{º}$.\label{fig:LocalDos}}
\end{figure}
The diagonalization of the secular equation gives the eigenstates
in the momentum basis, as well as the eigenvalues, so it is straightforward
to calculate the local density of states. The density plots shown
in Fig.~\ref{fig:LocalDos} show the local DOS in the superlattice
unit cell, integrated over a narrow energy range, close to zero; the
dispersioneless band is composed of states localized in the $AA$
stacking region, as was first found de Laissardière and co-workers
\cite{laissardiere2010}. 

\begin{figure}
\includegraphics[width=0.9\columnwidth]{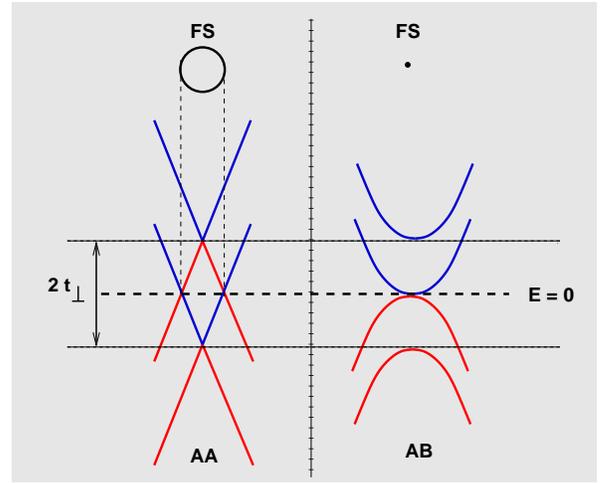}

\caption{Band structures of pure $AA$ and pure $AB$ stacking bilayers.\label{fig:AA-AB_interface}}

\end{figure}

The reason for this localization, and for the curious fact that the
degree of localization can oscillate with angle can be traced to the
difference of band structures of $AA$ and $AB$ (or $BA$) bilayers.
In a twisted bilayer of small angle, there are well defined regions
of $AA$, $AB$ and $BA$ stacking, and it is legitimate to reason
in terms of the corresponding band structures. The band structure
for an $AA$ stacked bilayer, near each Dirac point, is composed of
two cones shifted in energy by $\pm t_{\perp}$, corresponding to
the bonding and anti-bonding combinations of $p_{z}$ orbitals in
each plane (Fig.~\ref{fig:AA-AB_interface}). As a result, the Fermi
surfaces for electrons and holes, at zero energy, are circular with
radius $k_{F}=t_{\perp}/\hbar v_{F}$. But on an $AB$ or $BA$ stacked
bilayer at zero energy the Fermi surface is a point. From conservation
of momentum parallel to an $AA\leftrightarrow AB$ interface, one
can immediately conclude that there can be no transmission $AA\to AB(BA)$
for any nonzero angle; in fact, a calculation shows that the transmission
coefficient is also zero for zero angle of incidence and energy. Since
the $AA$ region is enclosed by an hexagon of $AB$ and $BA$ stacking,
this raises the possibility of localization of zero energy states
in the $AA$ region. However, this localization only occurs for zero
energy; for finite energy some transmission is possible. Now, a confined
$AA$ region will have a discrete spectrum. The energy levels move
toward the corresponding Dirac energies ($\pm t_{\perp})$ if the
size of the unit cell size increases. Is is clear then, that when
\emph{a discrete level in an $AA$ region occurs at zero energy},\emph{
}we can expect strong localization and a dispersioneless band. If
we further decrease the angle, by increasing the unit cell, the discrete
level moves away from zero energy, it starts tunneling into the neighboring
$AB$ and $BA$ regions, and the low energy band broadens. This, we
believe, is the rather simple explanation for the oscillation of the
bandwidth of the central peak in the DOS. An explanation for this
same oscillation, formulated in terms of non-Abelian effective gauge
fields, has recently been proposed\cite{Guinea2011}. The extremely
flat bands --- which Bistritzer and MacDonald \cite{ISI:000293129900015}
associate with the zeros of the Fermi velocity --- correspond to the
passage of a state of a confined $AA$ region through energies $\pm t_{\perp}$
above (below) the corresponding Dirac energies (\emph{i.e. }zero energy).
The double peak structures arises from the presence of electron and
hole states at zero energy in the $AA$ regions.

\section{Discussion and Conclusions}

We have analyzed in detail the continuum description of the twisted
bilayer focusing on small angle structures. We generalized our previous
treatment to include all types of commensurate structures, and addressed
in particular the possibility of a gapped electronic spectrum for
SE-even structures raised by Mele\cite{PhysRevB.81.161405}. We have
shown, that for small angles, all commensurate structures are either
of the type $r=1$, in which the relation between the period and angle
or rotation is that found in STM studies of Moire patterns, $L=a/\left[2\sin(\theta/2)\right]$,
or \emph{almost periodic }repetitions of such structures. As a consequence,
even though the momentum space description can be quite different,
small angle commensurate structures share the same physics. 

We have achieved a complete analytical characterization of the Fourier
components of the spatially modulated hopping amplitudes, which allows
a detailed study of very small angle structures. This continuum description
accounts very well for the renormalization of the Fermi velocity relative
to the single layer value. The density of states is a revealing tool;
if the angle is not two small, two well defined Van-Hove peaks appear
at low energies, and, near zero energy, the DOS rises linearly, as
expected for linear dispersion (Fig.~\ref{fig:DOSm18}); at $\theta=1.08\text{º }$,
the Van-Hove peaks are no longer resolved, as the range of linear
dispersion shrinks to zero; a low energy, almost flat, band appears,
separated by gaps from the rest of the spectrum (at positive and negative
energies). This flat band is formed from states localized in $AA$
stacking regions which, at zero energy, cannot tunnel into $AB$ and
$BA$ regions. However, if the angle is further decreased the energy
of these localized states changes, and they can start tunneling into
the neighboring regions. This explains the oscillation with angle
of the bandwidth of the central peak of the density of states.
\begin{acknowledgments}
J. M. B. L. S. was supported by Fundação para a Ciência e a Tecnologia
(FCT) and is thankful for the hospitality of Boston University and
of National University of Singapore. N. M. R. P. was supported by
Fundos FEDER through the Programa Operacional Factores de Competitividade
- COMPETE and by FCT under project no. PEst-C/FIS/UI0607/2011, and
is thankful for the hospitality of Boston University. AHCN acknowledges
DOE grant DE-FG02-08ER46512, ONR grant MURI N00014-09-1-1063, and
the NRF-CRP award \textquotedbl{}Novel 2D materials with tailored
properties: beyond graphene\textquotedbl{} (R-144-000-295-281).
\end{acknowledgments}
\appendix

\section{Geometry\label{sec:Geometry}}

Given an arbitrary site of the hexagonal Bravais lattice, $\mathbf{P}_{1}=k\mathbf{a}_{1}+l\mathbf{a}_{2}$,
the rotational/reflection symmetry  implies that it is part of a set
of twelve $\{\mathbf{P}_{i},\mathbf{Q}_{i}:\, i=1,\dots,6\}$, $\pi/3$
being the angle between directions of consecutive points in $\{\mathbf{P}_{i}\}$
or in $\{\mathbf{Q}_{i}\}$, and each of these sets being the image
of the other under reflection about the symmetry axes (see Fig.~\ref{fig:A-shell-12}).
These two sets merge into one if and only if either $k$ or $l$ is
zero or $k=l$. In the main text, we argued that we need only consider
rotations that map one of these sets onto its image by reflection,
in order to obtain all angles and primitive vectors of commensurate
structures. Without loss of generality we can choose\begin{subequations}\label{P1_Q1_Q6}
\begin{eqnarray}
\mathbf{P}_{1} & = & n\mathbf{a}_{1}+m\mathbf{a}_{2}\label{eq:P1}\\
\mathbf{Q}_{1} & = & m\mathbf{a}_{1}+n\mathbf{a}_{2}\label{eq:Q1}\\
\mathbf{Q}_{6} & = & (m+n)\mathbf{a}_{1}-m\mathbf{a}_{2}.\label{eq:Q6}
\end{eqnarray}
 \end{subequations} with $n>m>0$; values of $m$ or $n$ zero, or
$m=n$, correspond to $\pi/3$ rotations, that transform an $AB$
stacked bilayer into an $AA$ one (see also Fig.~\ref{fig:A-shell-12}).
So anticlockwise commensurate rotations with angles $0<\theta<\pi/3$,
are of two types\begin{subequations}\label{eq:rots_12} 
\begin{eqnarray}
\theta:\qquad\mathbf{P}_{1} & \to & \mathbf{Q}_{1};\quad(n,m)\to(m,n);\label{eq:rot1}\\
\theta':\mathbf{\qquad Q'}_{6} & \to & \mathbf{P'}_{1};\quad(p+q,-p)\to(q,p);\label{eq:rot2}
\end{eqnarray}
\end{subequations}In the first case $\mathbf{T}_{1}:=m\mathbf{a}_{1}+n\mathbf{a}_{2}$
is a super-lattice translation; in the second it is $\mathbf{T}'=q\mathbf{a}_{1}+p\mathbf{a}_{2}$.
We will soon see under what conditions these are primitive vectors.
These two rotations are conjugate, $\theta+\theta'=\pi/3$, if $m=p$
and $n=q$.

In the following, it will be useful to to define these rotations in
terms of the pair of integers $m,r$ with $r=n-m$, and $p,s$ with
$s=q-p$:\begin{subequations}\label{eq:two_rots2}
\begin{eqnarray}
(m+r,m) & \to & (m,m+r)\label{eq:conjugate_m_r_1}\\
(2p+s,-p) & \to & (p+s,p).\label{eq:conjugate_m_r2}
\end{eqnarray}
\end{subequations}One easily derives the following results for the
angles, by taking the scalar product of final and initial vectors\begin{subequations}\label{eq:cos}
\begin{eqnarray}
\cos\theta & = & \frac{3m^{2}+3mr+r^{2}/2}{3m^{2}+3mr+r^{2}}\label{eq:cos_typeI}\\
 &  & =\frac{3(m+r/2)^{2}-(r/2)^{2}}{3(m+r/2)^{2}+(r/2)^{2}}\\
\cos\theta' & = & \frac{3p^{2}/2+3ps+s^{2}}{3p^{2}+3ps+s^{2}}\label{eq:cos_typeII}\\
 &  & =\frac{3(3p/2+s)^{2}-(3p/2)^{2}}{3(3p/2+s)^{2}+(3p/2)^{2}}
\end{eqnarray}
\end{subequations}The second form of each expression makes it clear
that the two families define the same set of angles: $\theta=\theta'$,
if $m/r=s/3p$; all angles of commensurate structures are generated
Eq.~(\ref{eq:cos_typeI}) with $m$ and $r$ positive integers: $\theta'(p,s)=\theta(m,r)$
if $m=s$ and $r=3p$.

Given two positive integers, $m,r$, and the angle $\theta(m,r)$
defined by Eq.(\ref{eq:cos_typeI}), there is a unique set of integers
$p,q$ for which one of the following representations \begin{subequations}\label{eq:angles}
\begin{eqnarray}
\cos\theta(m,r) & = & \frac{3p^{2}+3pq+q^{2}/2}{3p^{2}+3pq+q^{2}},\label{eq:angle_typeI}\\
\cos\theta(m,r) & = & \frac{3p^{2}/2+3pq+q^{2}}{3p^{2}+3pq+q^{2}}\label{eq:angle_typeII}
\end{eqnarray}
\end{subequations}has the \emph{smallest denominator.} If\emph{ }the
smallest denominator occurs for the first form, we conclude that $\mathbf{t}_{1}:=p\mathbf{a}_{1}+(p+q)\mathbf{a}_{2}$
is a lattice translation, with the smallest norm (the denominator
is $\left|\mathbf{t}_{1}\right|^{2}$) and, therefore, a primitive
vector. The other can be obtained by a $\pi/3$ rotation of $\mathbf{t}_{1}$.
On the other hand, if the second form has the smallest denominator,
then, by the same reasoning, $\mathbf{t}_{1}=(p+q)\mathbf{a}_{1}+p\mathbf{a}_{2}$
is a primitive vector of the super-lattice. 

From this point on, we assume that $m,r$ are co-prime, because otherwise
we can always reduce the denominator by factoring out the divisors
of $m$ and $r$. If

\begin{equation}
\frac{3m^{2}+3mr+r^{2}/2}{3m^{2}+3mr+r^{2}}=\frac{3p^{2}+3pq+q^{2}/2}{3p^{2}+3pq+q^{2}},\label{eq:two_fracs_mr_pq}
\end{equation}
and $3p^{2}+3pq+q^{2}<3m^{2}+3mr+r^{2}$, we must have,\begin{subequations}
\begin{eqnarray}
3m^{2}+3mr+r^{2}/2 & = & \lambda\left(3p^{2}+3pq+q^{2}/2\right),\\
3m^{2}+3mr+r^{2} & = & \lambda\left(3p^{2}+3pq+q^{2}\right),
\end{eqnarray}
\end{subequations}where $\lambda$ is a positive integer. Subtracting
these equations, one gets $r^{2}=\lambda q^{2}$, so that $\lambda=s^{2}$,
where $s$ is a divisor of $r$. Solving the second equation for $m/s$,
gives, recalling that $m,r,p,q$ are positive integers, 
\begin{eqnarray}
\frac{m}{s} & = & -\frac{q}{2}\pm\frac{1}{2}\sqrt{q^{2}+4p(p+q)}=p\label{eq:msp-1}
\end{eqnarray}
So $s$ must a common divisor of $m$ and $r$, and, since $m,r$
are co-prime, $s=1$, and the initial form already has the smallest
denominator. An entirely similar argument can applied to reducing
to the second form (Eq.~(\ref{eq:angle_typeII})). A form with smaller
denominator is possible if $r$ is a multiple of 3, and $(p,q)=(m,r/3)$.

In conclusion, we can state that if $(m,r)$ are co-prime and 
\[
\cos\theta=\frac{3m^{2}+3mr+r^{2}/2}{3m^{2}+3mr+r^{2}}
\]
the super-lattice basis vectors are given by $\mathbf{t}_{i}=\sum_{j}S_{ij}\mathbf{a}_{j}$,
and the matrix $\mathbf{S}$ is defined in Eqs.~(\ref{eq:primitive_vec_typeI})
and (\ref{eq:primitive_vecs_typeII-1}). Shallcross \emph{et. al.
}define the angles and primitives vectors in terms of two co-prime
integers $p$ and $q$; their results coincide with these with the
following correspondence: if $r$ is odd, $p=r$ and $q=2m+r$; is
$r$ is even, $p=r/2$ and $q=m+r/2$. 

From these results one can obtain other useful relations. Since $\mathbf{t}_{1}$
and $\mathbf{t}_{2}$ are lattice translations of \emph{both }layers
there have equally simple expressions in terms of the primitive vectors
of the rotated layer $\mathbf{a}_{1}'$ and $\mathbf{a}_{2}'.$ The
transformation between these non-orthogonal basis is 
\begin{equation}
\left[\begin{array}{c}
\mathbf{a}_{1}\\
\mathbf{a}_{2}
\end{array}\right]=\left[\begin{array}{cc}
\cos\theta+\sin\theta/\sqrt{3} & -2\sin\theta/\sqrt{3}\\
2\sin\theta/\sqrt{3} & \cos\theta-\sin\theta/\sqrt{3}
\end{array}\right]\left[\begin{array}{c}
\mathbf{a'}_{1}\\
\mathbf{a}'_{2}
\end{array}\right].\label{eq:rot-matrix_aps_to_as-1}
\end{equation}
The rotation matrix can be expressed in terms of $m$ and $r$ using
Eq.~(\ref{eq:cos_typeI}), leading to 
\begin{eqnarray}
\left[\begin{array}{c}
\mathbf{t}_{1}\\
\mathbf{t}_{2}
\end{array}\right] & = & \left[\begin{array}{cc}
m+r & m\\
-m & 2m+r
\end{array}\right]\left[\begin{array}{c}
\mathbf{a'}_{1}\\
\mathbf{a}'_{2}
\end{array}\right]\label{eq:aprimes_to_ts_typeI}
\end{eqnarray}
 for $\gcd(r,3)=1$ and 
\begin{eqnarray}
\left[\begin{array}{c}
\mathbf{t}_{1}\\
\mathbf{t}_{2}
\end{array}\right] & = & \left[\begin{array}{cc}
m+2r/3 & -r/3\\
r/3 & m+r/3
\end{array}\right].\left[\begin{array}{c}
\mathbf{a'}_{1}\\
\mathbf{a}'_{2}
\end{array}\right]\label{eq:aprimes_to_ts-typeII}
\end{eqnarray}
for $\gcd(r,3)=3$. 

The dual basis of $\{\mathbf{a}_{1},\mathbf{a}_{2}\}$ (reciprocal
lattice primitive vectors) can be chosen as 
\begin{equation}
\left[\begin{array}{c}
\mathbf{g}_{1}\\
\mathbf{g}_{2}
\end{array}\right]=\frac{4\pi}{3\left|\mathbf{a}_{1}\right|}\left[\begin{array}{cc}
2 & -1\\
-1 & 2
\end{array}\right]\left[\begin{array}{c}
\mathbf{a}_{1}\\
\mathbf{a}{}_{2}
\end{array}\right],\label{eq:dual_of_a1_a2}
\end{equation}
with a similar relation for $\{\mathbf{t}_{1},\mathbf{t}_{2}\}$ and
its dual basis $\{\mathbf{G}_{1},\mathbf{G}_{2}\}$. Knowing that
the Dirac points are given as $\mathbf{K}=(4\pi/3)(\mathbf{a}_{1}-\mathbf{a}_{2}),$
and $\mathbf{K}^{\theta}=(4\pi/3)(\mathbf{a}'_{1}-\mathbf{a}'_{2})$,
one can show, using Eqs.~(\ref{eq:rot-matrix_aps_to_as-1}) to (\ref{eq:dual_of_a1_a2})
and Eqs.~(\ref{eq:primitive_vec_typeI},\ref{eq:primitive_vecs_typeII-1}),
after some tedious but trivial algebra, the following relations,
\begin{equation}
\Delta\mathbf{K}^{\theta}:=\mathbf{K}^{\theta}-\mathbf{K}=\begin{cases}
\frac{r}{3}\left(2\mathbf{G}_{1}+\mathbf{G}_{2}\right) & \textrm{if }\gcd(r,3)=1\\
\frac{r}{3}\left(\mathbf{G}_{1}+\mathbf{G}_{2}\right) & \textrm{if }\gcd(r,3)=3.
\end{cases}\label{eq:delta_K}
\end{equation}
Note that, in the second case only, $\Delta\mathbf{K}$ is a reciprocal
lattice vector.


\end{document}